%% file: main_text_rev1.tex
\documentclass[
  aps,
  pre,
  reprint,
  superscriptaddress,
  floatfix
]{revtex4-2}

\input{operands}

\usepackage{multirow}
\usepackage{setspace}
\usepackage{enumitem}
\usepackage{verbatim}

\usepackage{amssymb}
\usepackage{amsbsy}
\usepackage[T1]{fontenc}
\usepackage{amsmath}
\usepackage{mathtools}

\usepackage{graphicx}
\usepackage{xcolor}
\usepackage{mathrsfs}
\usepackage[mathscr]{euscript}
\usepackage{tikz}
\usetikzlibrary{shapes}
\usepackage{cancel}
\usepackage[normalem]{ulem}

\usepackage[colorinlistoftodos]{todonotes}
\usepackage{verbatim}
\usepackage{latexsym}
\usepackage{textcomp}
\usepackage{bm}
\usepackage[Euler]{upgreek}
\usepackage{longtable}
\usepackage{subfigure}

\usepackage[colorlinks,allcolors=cyan!70!black]{hyperref}

\usepackage{epsfig}
\usepackage{dcolumn}

\usepackage{booktabs}

\newcommand{\qed}{\nobreak \ifvmode \relax \else
      \ifdim\lastskip<1.5em \hskip-\lastskip
      \hskip1.5em plus0em minus0.5em \fi \nobreak
      \vrule height0.75em width0.5em depth0.25em\fi}

\DeclareMathAlphabet\mathbfcal{OMS}{cmsy}{b}{n}
\DeclareMathAlphabet{\mathbfsf}{\encodingdefault}{\sfdefault}{bx}{sl}

\begin{document}

\setlength{\parindent}{1em}
\setlength{\parskip}{0pt}

\setlength{\abovedisplayskip}{6pt}
\setlength{\belowdisplayskip}{6pt}
\setlength{\abovedisplayshortskip}{4pt}
\setlength{\belowdisplayshortskip}{4pt}

% ---- your paper starts here ----

%\preprint{}

\title{Rigidity transition in polydisperse shear-thickening suspensions}


\author{Sourav Kumar Singh}
\affiliation{Department of Chemical Engineering, Indian Institute of Technology (ISM) Dhanbad, Jharkhand 826004, India}
\author{Vishant Tyagi}
\affiliation{Department of Chemical Engineering, Indian Institute of Technology (ISM) Dhanbad, Jharkhand 826004, India}
\author{Aritra Santra}
\email{aritrasantra@iitism.ac.in}
\affiliation{Department of Chemical Engineering, Indian Institute of Technology (ISM) Dhanbad, Jharkhand 826004, India}
 

\date{\today}

%\vspace{-20pt}

%\begin{figure}[ptbh]
% \begin{center}
% {\includegraphics*[width=!,height=7.5cm]{fig_TOC.pdf}}
% \end{center}
% \vskip-10pt
% \text{For Table of Contents use only}
%\end{figure}

%----------------------------------------------------------------------------------
\begin{abstract}
%----------------------------------------------------------------------------------
Shear thickening suspensions of non-Brownian polydisperse particles are simulated in 2D using a discrete element method based algorithm (LF-DEM) at high packing fractions ($\phi$) and large non-dimensional stresses ($\sigma$). Rigidity analysis of the stress induced particle clusters is carried out using \textit{pebble game} algorithm for polydisperse suspensions and compared with the statistically equivalent bidisperse systems. A critical value of the packing fraction, $\phi_c$, close to the shear jamming transition, $\phi_J^{\mu}$, ($\phi_c<\phi_J^{\mu}$) is obtained where rigid particle clusters begin to grow sharply. The growth is found to be characterized by a critical transition of an order parameter ($f_\text{rig}$), defined by the fraction of particles in rigid clusters which scales as, $f_\text{rig}\sim (\phi-\phi_c)^{\beta}$ for $\phi>\phi_c$, and by the susceptibility scaling, $\chi_\text{rig}\sim|\phi-\phi_c |^{-\gamma}$, {\color{black}with exponents having values consistent with the critical exponents in 2D percolation transition}. The variations of $f_\text{rig}$ and $\chi_\text{rig}$ in polydisperse suspensions are found to be identical to that of the statistically equivalent bidisperse suspensions. {\color{black}Finite size scaling analysis shows a divergence of correlation length near $\phi_{c_\infty}$ following critical exponent $\nu\approx 1.33$, in agreement with the 2D percolation theory}. Furthermore, $\phi_c$ and $\phi_J^{\mu}$ are found to vary non-monotonically with polydispersity index and depend on the particle stiffness.
%----------------------------------------------------------------------------------
\end{abstract}
%\keywords{}

\maketitle

%\section{\label{sec:intro}Introduction}

%----------------------------------------------------------------------------------
\section{\label{sec:intro} Introduction}
%----------------------------------------------------------------------------------
The flow behavior of dense suspensions of non-Brownian particles has attracted the attention of researchers for the past several decades because of the rich physics involved in these apparently simple systems and for their wide range of industrial applications such as, in chocolate refining, concrete mixing, ceramic processing, rocket fuel, etc~\cite{Brown_Jaeger2014,Ness2022,Blanco2019,Feys2009}. It is well known that in shear flow, depending on the particle volume fraction, applied stress, and particle interaction forces, dense suspensions show rate-dependent rheology such as, continuous shear thickening (CST) and discontinuous (DST) shear thickening~\cite{Mari2014, Wyart_Cates2014, MariPRE2015,Morris2020,Denn_Morris2014}. Furthermore, depending on the strength of the externally applied shear stress ($\hat{\sigma}$) and frictional contact force between the particles (with $\mu$ being the static friction coefficient), dense suspensions undergo shear jamming at a solid volume fraction ($\phi_J^{\mu}$) much lower than the maximum random close packing fraction or frictionless jamming fraction ($\phi_J^{0}$)~\cite{Brown_Jaeger2014,Mari2014,Morris2020,Pradeep_PRL2021}. At sufficiently high solid fraction, close to shear jamming, the transition from CST to DST is explained by the increase in the number of stress-induced frictional contacts~\cite{SetoPRL2013,Wyart_Cates2014,Lin2015}. Typically, there exists a characteristic stress $\sigma_0 \approx F_0/a^2$, proportional to the repulsive force scale $F_0$ acting between particles of characteristic radius $a$, which determines the relative strength of the externally applied stress and controls the transition from frictionless ($\phi_J^{0}$) to frictional jamming ($\phi_J^{\mu}$)~\cite{Wyart_Cates2014,Mari2014,ASingh2018}. In the limit of low stress ($\hat{\sigma} \ll \sigma_0$), the contacts between the particles are mostly dominated by lubrication interaction leading to frictionless jamming at sufficiently high particle volume fraction. Whereas, in the high stress limit ($\hat{\sigma} \gg \sigma_0$), lubrication between the particles is overpowered by the frictional contact leading to frictional jamming. This transition in the flow property has been described by a cross-over scaling which correlates the increase in frictional contacts with increasing applied stress~\cite{NelyaAritra22,Meera_JOR2023,Wyart_Cates2014}. Moreover, the flow state of a shear thickening dense suspension depends on the applied stress, solid volume fraction and strength of frictional interactions. 

In addition to the applied stress and strength of frictional contacts, the size distribution of the particles immensely influence the shear thickening and jamming transition in these suspensions~\cite{Farris_1968,Guy2020,Nelya2023,AbhiSingh_PRE2024,Chang_JOR1994,Gondret_JOR1997,Yankai_2021}. For instance, at a fixed solid volume fraction $\phi$, bidisperse suspensions of any given ratio of large particle volume ($\phi_{\ell}$) to total particle volume fraction ($\zeta=\phi_{\ell}/\phi$), are found to show lower viscosity as compared to a mono-disperse suspension at the same condition~\cite{Chang_JOR1994,Pednekar2018,Nelya2023}. Furthermore, at a constant $\zeta$, the viscosity is observed to monotonically decrease with increasing bidispersity ratio ($\delta=a_{\ell}/a_s$), where $a_{\ell}$ and $a_s$ are the large and small particle size, respectively~\cite{Farris_1968,Pednekar2018}. This indicates more efficient packing of the particles in bidisperse suspensions as compared to monodisperse systems. Similar trends were observed experimentally for bidisperse suspensions of polymethylmethacrylate (PMMA) particles~\cite{Guy2020}. However, the Wyart and Cates (WC) model~\cite{Wyart_Cates2014}, used to explain the CST and DST in dense suspensions, failed to predict the shear thickening behavior for PMMA suspensions with a predominant fraction of large particles~\cite{Guy2020}. The results suggest that the underlying physics in the shear thickening behavior of dense suspensions also depends on the type of contacts between the particles, i.e. whether large-large, small-small or large-small contact is predominating. Recently, \citet{Nelya2023} simulated shear thickening suspensions with different values of bidispersity ratio ($\delta$) over a range of shear stress and observed the CST-DST transition to be dependent on the composition of small and large particles, supporting the observations by \citet{Guy2020}. Nevertheless, they show large particle ordering in the low stress limit to explain a reduction in viscosity at high value of $\zeta (=0.85)$ and found that the Maron-Pierce (M-P) model~\cite{Maron1956} can fit the viscosity divergence with packing fraction both in the low and high stress limits~\cite{Nelya2023}. Another important feature in bidisperse shear thickening suspensions is the non-monotonicity of the relative suspension viscosity ($\eta_r$) with $\zeta$ at a fixed bidispersity ratio $\delta$~\cite{Pednekar2018,Nelya2023,Guy2020,Chang_JOR1994}. Several models have been proposed to explain this behavior by correlating the maximum packing fraction $\phi_J$ with the particle size distribution~\cite{Qi_Tanner_KARJ2011,Qi_Tanner_RheoActa2012,Carmine_JCP2023,Pishvaei_CES2006}. More recently,~\citet{AbhiSingh_PRE2024} proposed a model based on a micro-mechanical picture which indicates that the non-monotonicity in the relative viscosity is related to the non-monotonic behavior of the stress contribution due to large-small particle contacts. Such a finding reiterates the important role of the types of particle contacts in affecting the rheology of bidisperse suspensions. While the bidisperse suspensions are the simplest form of polydisperse systems and are relatively easier to study, materials typically handled in industries and other practical problems generally consist of polydisperse particles with more than two particle sizes. However, due to the presence of multiple particle sizes, the effects of particle contact types on the rheology and flow microstructure are much difficult to analyse statistically. Interestingly, it has been shown that the rheology of a polydisperse suspension, in terms of viscosity, normal stresses, particle pressures and packing efficiency of the particles, is identical to that of a statistically equivalent bidisperse suspension~\cite{Desmond_2014,Ogarko_SM2013,Pednekar2018,Gu_PowTec2016}. In case of dry granular materials, similar analogy between the flow behavior of monodisperse and polydisperse systems is established by matching the grain size distribution~\cite{Polania_PRE2025}. However, while the bulk rheological properties of monodisperse and polydisperse shear thickening suspensions are well studied, a comprehensive understanding of the microstructural features leading to CST-DST transition and shear jamming in these systems is still missing. Specifically, the mechanisms involved in the growth of flow microstructure leading to a shear jammed state with increasing solid fraction, and how it is affected by particle size distribution is not well understood. Numerical analysis of the frictional contact networks and cluster rigidity suggests a correlation between huge stress fluctuations and percolation transition at the onset of DST and shear-jamming~\cite{Sedes_JoR2020,Sedes2022,Goyal_JOR2024,Mike2024}. Shear jamming in the high stress limit is found to be preceded by a non-equilibrium critical phenomenon described by a rigidity transition of frictionally contacted particle clusters~\cite{Santra_PRR2025}. While these studies are focused on understanding the correlation between the growth of the frictional contact network and DST or onset of shear jamming in nearly mono-disperse (or bidisperse suspensions with small $\delta$) suspensions, effects of polydispersity on the growth of flow microstructure is not well explored. In this work, we have studied the growth and transition of cluster rigidity near shear jamming in the high stress limit for systems of polydisperse suspensions.

Recently, \citet{Mike2024} and \citet{Santra_PRR2025} have implemented a constraint counting algorithm (\textit{pebble game})~\cite{Jacobs1997,Silke2016,Liu_Henkes_PRX2019} to compute fraction of rigid clusters in two dimensional (2D) bidisperse suspensions to understand the underlying mechanism in shear jamming at high stresses. Notably,~\citet{Santra_PRR2025} reported a critical phenomenon associated with the shear jamming at large stresses based on an order parameter computed from the rigid particle fraction, with critical exponents similar to the 2D-Ising model. Inspired from these studies, here we have investigated the effects of particle size distribution on the formation of rigid clusters and the associated critical transition near shear jamming at large stress ($\hat{\sigma}\gg \sigma_0$). Specifically, we compare the viscosity, flow microstructure and gradient direction velocity correlation functions of polydisperse systems with the statistically equivalent bidisperse suspensions. Moreover, our analysis demonstrates that shear jamming in polydisperse suspensions is also associated with a critical phenomenon which is identical to the statistically equivalent bidisperse systems. {\color{black}We also perform finite size scaling to characterize the divergence of the correlation length and susceptibility near the critical transition and compare the critical exponents to that of 2D percolation transition.} Broadly, we divided the rest of the paper into the following sections. Section~\ref{sec:GovEq} discusses the modeling and simulation methodology. The method of analysis and validation study for polydisperse systems are presented in section~\ref{sec:Valid}. The key results are discussed in section~\ref{sec:Results} and the main conclusions are highlighted in section~\ref{sec:Conclusions}.   

%%%%%%%%%%%%%%%%%%%%%%%%%%%%%%%%%%%%%%%%%
\section{Modeling and simulations}\label{sec:GovEq}
%%%%%%%%%%%%%%%%%%%%%%%%%%%%%%%%%%%%%%%%%%%%%%%%%%%%%%%%%%%%%%%%%%%%%

In the present study we have simulated systems of {\color{black}highly concentrated polydisperse suspensions of non-Brownian particles in 2D, considering a monolayer of $N$ spherical particles confined on a plane in a simulation box. We have adapted this model for simulating 2D suspensions from the work by~\citet{Seto2013JOR}. It is well established in the literature that the flow curves of shear thickening suspensions simulated in 2D and 3D are qualitatively similar~\cite{Gameiro2020,Nabizadeh2022,Santra_PRR2025}. In fact, the flow curves for 2D and 3D systems show nearly quantitative agreement if the packing fraction $\phi$ is appropriately mapped.} However, we have chosen a 2D system for this study to implement the pebble game algorithm for the analysis of rigid clusters, which is only applicable in 2D as discussed in the subsequent sections. {\color{black}We agree that in 3D system the types of cluster would be topologically different from that observed in 2D, however, that does not result in any qualitative difference in the macroscopic flow properties. Thus 2D suspensions still show all the qualitative features of the flow behavior similar to 3D systems.} The simulations are carried out using the well-established Lubrication Flow - Discrete Element Method (LF-DEM) algorithm developed by~\citet{Mari2014,SetoPRL2013}. The simulation methodology involved modeling of neutrally buoyant non-Brownian spherical particles with specified particle size distribution suspended in a Newtonian solvent, where the particles interact via pairwise lubrication interaction, frictional contact force and short range electrostatic repulsion. Simple shear flow is implemented by using Lees-Edward periodic boundary conditions. {\color{black} The LF-DEM simulation method is developed considering the flow to be in the regime of low Stokes number ($St\ll 1$) and particle Reynolds number, thereby neglecting the inertial term in the equations of motion. Furthermore, the size of the model particles are considered to be in the micron range where Brownian forces are unimportant. The long-range hydrodynamic interactions between the particles are assumed to be screened since the systems are highly concentrated and therefore, only pair-wise lubrication interactions are considered. In our simulations we have considered a constant microscopic friction coefficient where frictional states are only dependent of the applied stress. It is appropriate to mention here that while shear banding in concentrated dry granular media, which is a system very similar to the dense suspensions except for the hydrodynamic part, is an important physical phenomenon, in our study we do not observe it. This aspect is well studied in the literature and it has been found that for suspensions of non-Brownian particles with fixed microscopic friction, shear banding is generally not observed~\cite{Lematre2009,Boyer2011,Hermes2016}. Moreover, LF-DEM provides an efficient and simplistic model for simulating systems of non-Brownian suspensions in the limit of high concentration. However, it is to be noted that the method will not work for dilute or semi-dilute suspensions where long-range hydrodynamic interactions become important. The method also requires significant modifications in case of simulating colloidal suspensions where Brownian forces are non-negligible. Considering all the assumptions discussed above, the equation of motion for each particle is given by the following force and torque balances derived from the Langevin equation in the overdamped limit, neglecting the Brownian forces.}

\begin{align}
\mathbf{F}_{h}\left(\mathbf{x},\mathbf{u}\right) + \mathbf{F}_{c}\left(\mathbf{x}\right) + \mathbf{F}_{r}\left(\mathbf{x}\right) = \mathbf{0}  \\
\mathbf{T}_{h}\left(\mathbf{x},\mathbf{u}\right) + \mathbf{T}_{c}\left(\mathbf{x}\right) = \mathbf{0} 
\end{align}

Here, $\mathbf{x}$ and $\mathbf{u}$ represent the position and velocity vectors for $N$ particles, respectively. The subscripts $h$, $c$ and $r$ denote the hydrodynamic, contact and repulsive components of the force or torque, respectively. The force and torque balance equations are solved for the translational and rotational components of the particle velocities, which is subsequently integrated to update the particle positions in each simulation step. It should be noted that the repulsive force does not generate any torque and is excluded from the torque balance equation. The hydrodynamic force is modeled as a sum of single-particle drag force and pair-wise lubrication interaction as shown below, 
\begin{align}
\mathbf{F}_{h} ={}&
-\mathbf{R}_{FU}(\mathbf{x})
\cdot \bigl[\mathbf{u}
- \mathbf{u}^{\infty}(\mathbf{x})\bigr] \nonumber\\
&+ \mathbf{R}_{FE}(\mathbf{x})
:\mathbf{E}^{\infty}
\end{align}

where, $\mathbf{E}^{\infty} = \dot{\gamma}(\hat{\mathbf{e}}_{x}\hat{\mathbf{e}}_{y}+\hat{\mathbf{e}}_{y}\hat{\mathbf{e}}_{x})/2$ is the rate of deformation tensor, $\mathbf{u}^{\infty} = \dot{\gamma}\,y\,\mathbf{e}_{x}$ is the imposed flow field with $\dot{\gamma}$ as the shear rate, and $x$ and $y$ are the flow and flow gradient directions, respectively. Considering the \textit{squeeze, shear} and \textit{pump} modes of lubrication interactions, $\mathbf{R}_{FU}\left(\mathbf{x}\right)$ and $\mathbf{R}_{FE}\left(\mathbf{x}\right)$ are the position dependent resistance tensors giving the hydrodynamic forces from the velocity and deformation, respectively~\cite{Ball_Melrose1997}. For a pair of particles with size $a_i$ and $a_j$ at a non-dimensional interparticle distance of $h_{ij} = 2(r_{ij}-a_i-a_j)/(a_i+a_j)$, the squeeze mode of resistance is proportional to $1/(h+\delta)$, whereas the pump and shear modes are proportional to $\log(h+\delta)$. Here, we have used $\delta= 10^{-3}$ as a regularization parameter, which mimics the surface roughness of the particles~\cite{Mari2014}. Considering the screening of the long-range hydrodynamic interactions in a concentrated suspension, only short range lubrication is implemented in the present model. For frictional contacts, stick/slide friction model is implemented with linear springs and dashpots~\cite{Cundall_1979,Luding_2008}. The frictional contact force consists of normal $\mathbf{F}_{c,n}^{(i,j)}$ and tangential $\mathbf{F}_{c,\tan}^{(i,j)}$ components, which are given as follows,
\begin{eqnarray}
    \mathbf{F}_{c,n}^{(i,j)} &=& k_n h_{ij}\bm{n}_{ij} + \gamma_n\mathbf{U}_n^{(i,j)}  \\
    \mathbf{F}_{c,\tan}^{(i,j)} &=& k_t\,\xi^{(i,j)} 
\end{eqnarray}

where, $k_n$ and $k_t$ are the normal and tangential spring constants, respectively, with $k_t=0.9\,k_n$, $\gamma_n$ is the damping coefficient for the normal component, $\mathbf{U}_n^{(ij)}$ is the normal component of the relative velocity between $i^{th}$ and $j^{th}$ particles, and $\xi^{(ij)}$ is the tangential stretch. The non-sliding frictional contacts between the particles are satisfied by the Coulomb friction law $|\mathbf{F}_{c,\tan}^{(i,j)}|\leq\mu|\mathbf{F}_{c,n}^{(i,j)}|$ with an interparticle friction coefficient $\mu$. {\color{black}To mimic the stabilizing force observed in many experimental systems of non-Brownian suspensions, we have implemented a simplistic form of the electrostatic double-layer repulsive force given by,}

\begin{equation}
  \mathbf{F}_r^{(ij)}(h) =
    \begin{cases}
      -2F_0\frac{a_ia_j}{a\,(a_i+a_j)}\exp(-h/\lambda)\bm{n}_{ij} & \text{if $h_{ij}\ge0$,} \nonumber \\
      -2F_0\frac{a_ia_j}{a\,(a_i+a_j)}\bm{n}_{ij}  & \text{if $h_{ij}<0$,}  \label{eq:ESR}
    \end{cases}   
    \quad \tag{6}
\end{equation}

\setcounter{equation}{6}
 where $\lambda = 0.05\,a$ is considered as the Debye length. {\color{black}We would like to emphasize that we have used here an approximate form the double-layer electrostatic repulsion in the limit of small Debye length as compared to the particle radius~\cite{Israelachvili2011,Mari2014}. This approximation works well for the present study since the size of the non-Brownian particles are in the range of several microns and the corresponding Debye length $\lambda$ is of the order of $10^{-1}\, \mu m$. It should also be noted that the repulsive force is bounded to a maximum value since the particles are prevented from overlapping by the strong pairwise lubrication interaction with a regularization parameter $\delta$.} Further details on the simulation model are available in the work by~\citet{Mari2014}. 

The simulations are carried out by keeping the stress fixed, \emph{i.e.} in a stress-controlled mode~\cite{ASingh2018,Nelya2023}, such that the shear rate can be computed from the equation,
\begin{equation}
    \dot{\gamma} = \frac{\sigma-\sigma_c-\sigma_r}{\eta_0(1+2.5\,\phi)+\eta_H} 
\end{equation}
where, $\eta_0$ is the solvent viscosity, $\phi$ is the particle packing fraction, $\sigma_c$ and $\sigma_r$ are the shear stresses corresponding to the contact force and repulsive force, respectively. These stresses are computed as,
\begin{equation}
\sigma_{r,c} = V^{-1} ( \mathbf{x}\mathbf{F}_{r,c} - \mathbf{R}_{SU}\cdot\mathbf{R}^{-1}_{FU}\cdot\mathbf{F}_{r,c} )_{xy}  
\end{equation}

\noindent {\color{black}The hydrodynamic stress is evaluated from the following relation}
\begin{equation}
\dot{\gamma}\eta_H = V^{-1} [ (\mathbf{R}_{SE}-\mathbf{R}_{SU}\cdot\mathbf{R}^{-1}_{FU}\cdot\mathbf{R}_{FE}) : \mathbf{E}^{\infty} ]_{xy}
\end{equation}
\noindent where $V$ is the volume of the simulation domain, and the resistance matrices $\mathbf{R}_{SU}$ and $\mathbf{R}_{SE}$ determine the lubrication stresses from the particle velocity and rate of deformation, respectively. 

\begin{table}[ptbh]
\begin{center}
\begin{tabular}{ p{2.8cm} p{2.8cm} p{2.8cm} }
 \hline
 Quantity & dimensionless form & characteristic scale \\  
 \hline\hline
 particle radius ($\hat{a}_i$) & $a_i=\hat{a}_i/a$ & $a=1.29\,\ell_c$ \\ 
 \hline
 box size ($\hat{L}$) & $L=\hat{L}/a$ & $a=1.29\,\ell_c$ \\
 \hline
 stress ($\hat{\sigma}$) & $\sigma=\hat{\sigma}/\sigma_0$ & $\sigma_0=F_0/6\pi a^2$ \\
 \hline
 shear rate ($\dot{\gamma}$) & $\dot{\gamma}/\dot{\gamma}_0$ & $\dot{\gamma}_0=F_0/6\pi\eta_0a^2$ \\
 \hline
 reduced viscosity ($\eta_r$) & $\eta_r=\eta/\eta_0$ & solvent viscosity ($\eta_0$) \\
 \hline
 time ($\hat{t}$) & $t=\hat{t}\,\dot{\gamma}_0$ & $1/\dot{\gamma}_0$ \\
 \hline
 spring stiffness ($k_n$) & $k_n/6\pi\eta_0 a\dot{\gamma}$ & $6\pi\eta_0 a\dot{\gamma}$ \\
 \hline\hline
\end{tabular}
\vspace{3mm}
\caption{\small Definition of different parameters and quantities in reduced units. 
Note that $\ell_c \gg 1\,\mu$m is considered as the characteristic length unit.}
\label{tab:tab1}
\end{center}
\end{table}

All the simulations, if not mentioned otherwise, are carried out with $N=1000$ spherical particles, interacting with static friction coefficient $\mu=0.5$ at contact. {\color{black} The parameters and quantities used in the present study are all defined in terms of reduced units and a summary of this is provided in Table~\ref{tab:tab1}. It is to be noted that $\ell_c$, $F_0$ and $\eta_0$ are the fundamental quantities in our simulations and all other parameters, expressed in reduced units, are defined in terms of these fundamental quantities. In the present work, the quantity $\ell_c \, \gg 1\, \mu m$ is considered as the fundamental length unit and all the particle radii defined in terms of $\ell_c$ could be assumed to be well above $1\,\mu m$, resulting the suspensions to be non-Brownian.} We have studied the systems in the limit of high non-dimensional stress ($\sigma = \hat{\sigma}/\sigma_0$) values of 25 and 100, where shear jamming is nearly independent of the applied stress~\cite{Meera_JOR2023,ASingh2018}. For consistency, all the length scales in the simulations are non-dimensionalized by a characteristic particle radius $a=1.29\,\ell_c$. To simulate polydisperse suspensions we have considered normal and log-normal particle size distributions. The mean ($\langle a\rangle$), standard deviation ($s.d.$) and skewness ($s$) of the distributions are defined as follows,
\begin{eqnarray}
 \langle a\rangle &=& \displaystyle\frac{1}{N}\sum\limits_{i}a_i \\
 s.d. &=& \left(\displaystyle\frac{1}{N}\sum\limits_{i}a_i^2-\langle a\rangle^2\right)^{1/2} \\
s &=& \left(\displaystyle\frac{1}{N}\sum\limits_{i}(a_i-\langle a\rangle)^3\right)/s.d.^3
\end{eqnarray}
Typically, we refer to polydisperse system by its polydispersity index $\alpha$, defined as $\alpha=s.d./\langle a\rangle$. The details of the particle size distributions simulated in the present study is tabulated in Table~\ref{tab:tab2}.

\begin{table}[ptbh]
\begin{center}
\begin{tabular}{ p{3cm} p{2cm} p{2cm} p{1cm}}
 \hline
 Type & $\langle a\rangle$ & $s.d.$ & $s$ \\  
 \hline 
 \hline
 normal & 1.5 & 0.108 & 0 \\ 
 \hline
 normal & 1.5 & 0.2 & 0 \\
 \hline
 normal & 1.5 & 0.28 & 0 \\ 
  \hline
 normal & 1.5 & 0.4 & 0 \\
  \hline
 normal & 1.5 & 0.48 & 0 \\  
 \hline
log-normal & 1.197 & 0.448 & 0.694 \\ 
  \hline
 \hline
\end{tabular}
\vspace{5mm}
\caption {\small{Specifications of polydisperse suspensions. $\langle a\rangle$, $s.d.$ and $s$ denotes the mean, standard deviation and skewness of the distribution, respectively.}\label{tab:tab2}}
\end{center}
\end{table}

%%%%%%%%%%%%%%%%%%%%%%%%%%%%%%%%%%%%%%%%%
\section{Rigidity analysis and validation study}\label{sec:Valid}
%%%%%%%%%%%%%%%%%%%%%%%%%%%%%%%%%%%%%%%%%%%%%%%%%%%%%%%%%%%%%%%%%%%%%

\begin{figure*}[ptbh]
    \centerline{
    \begin{tabular}{c @{\hspace{18mm}} c}
        \includegraphics[width=60mm]{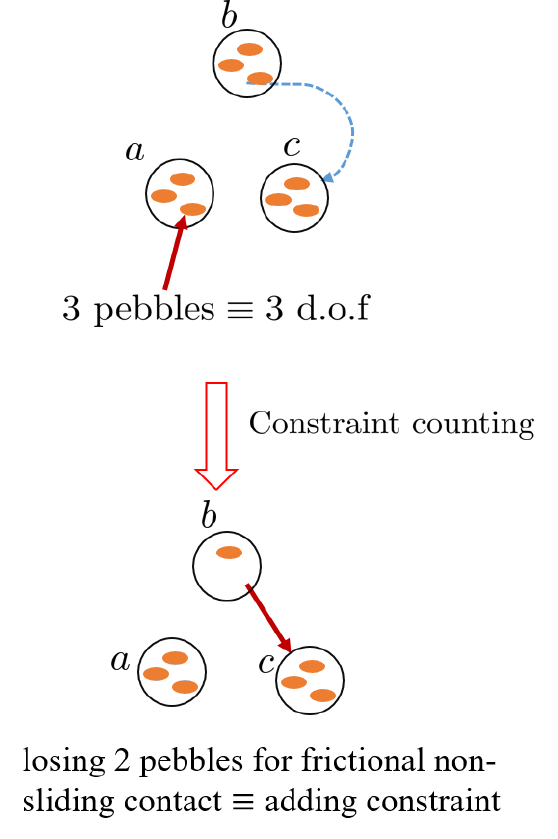} 
        & \includegraphics[width=80mm]{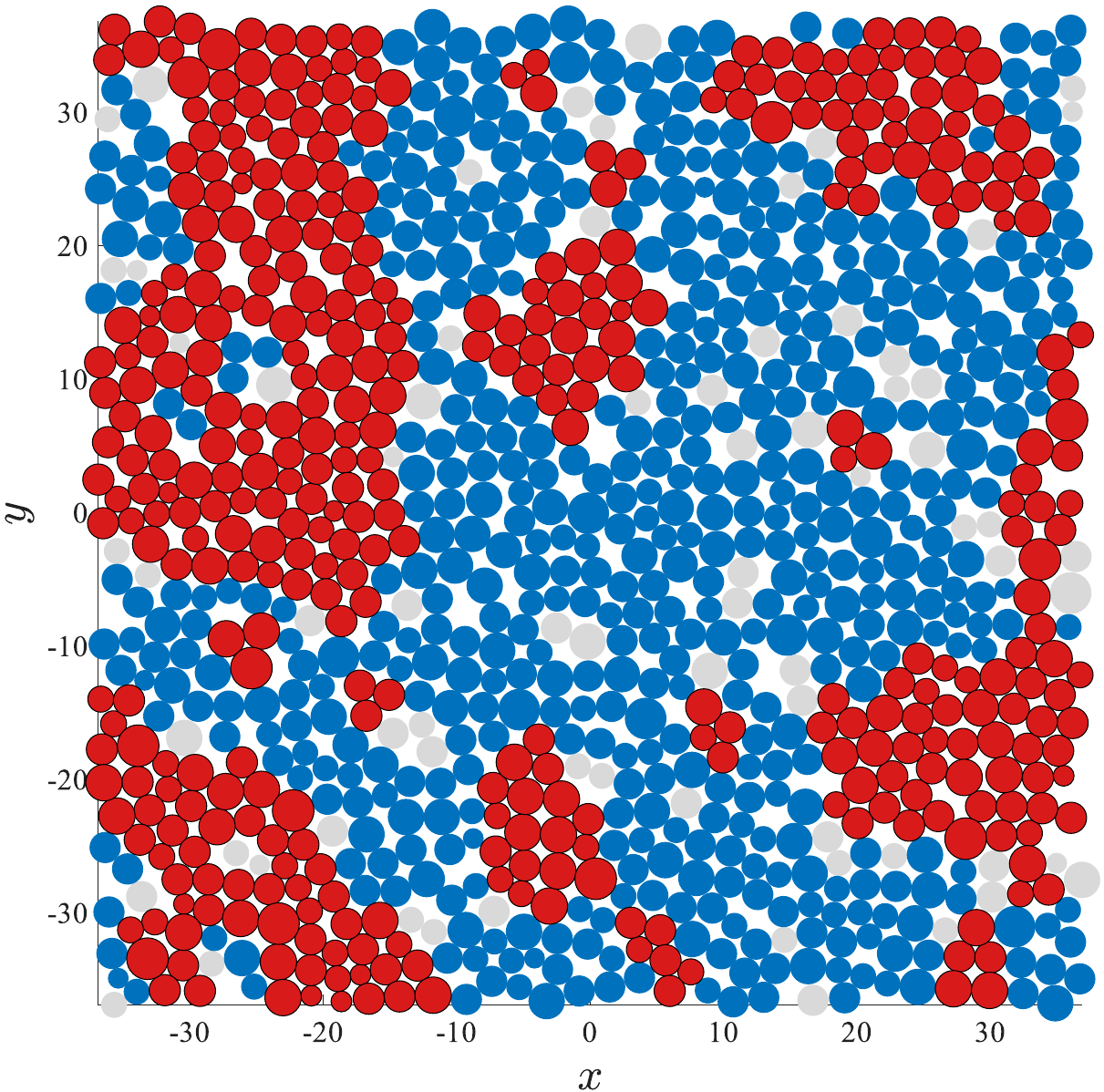} \\
        (a) & (b) \\
    \end{tabular}
    }
    
   \caption{\small{{\color{black}(a) Schematic diagram illustrating a single step of the (3,3) PG algorithm. (b) Simulation snapshot showing regions of rigid clusters identified by PG algorithm (indicated by red coloured particles), surrounded by non-rigid regions (with blue coloured regions representing particles in frictional contacts and grey zones representing particles which are interacting via non-frictional interactions).}} 
\label{fig:PG_snapshot}}
\end{figure*}

The identification of the rigid clusters from a given simulation snapshot is performed by applying the (3,3) pebble game (PG) algorithm, which is a constraint counting algorithm to identify the rigid nodes from a given graph~\cite{Jacobs1997,Maxwell_1864}. Here, each particle is assigned with 3 pebbles corresponding to 2 translational and 1 rotational degrees of freedom. In a contact network, whenever a pair of particles are in frictional contact, one of the particle loses one or two pebbles, depending on whether the contact is sliding or non-sliding, respectively, resulting in addition of constraints. {\color{black}We have illustrated this algorithm schematically in Fig.~\ref{fig:PG_snapshot}~(a)}. Likewise, for a given cluster of particles, all such constraints are constructed and the number of free pebbles are counted. If only a total of three free pebbles remain for the cluster, then the cluster is designated as a rigid one according to PG algorithm. {\color{black}It is to be noted that the rigid clusters are essentially transient structures identified by the PG algorithm, purely based on geometric criterion. The constraints implemented between the nodes (\emph{i.e.} between the particles) are only dependent on the states of frictional contact, where frictionless hydrodynamic and repulsive interactions are not accounted for. Thus, by rigidity we imply the mechanical stability of a sub-structure of frictional contact network surrounded by non-rigid flowable regions, which are continuously reforming the clusters. This is illustrated schematically by presenting a simulation snapshot in Fig.~\ref{fig:PG_snapshot}~(b), where we have highlighted regions of rigid clusters (represented by red colored particles) surrounded by non-rigid regions (represented by blue and gray particles).} Identification and growth of such rigid clusters have been considered in characterizing jamming in dry granular media~\cite{Silke2016,Ellenbroek_PRL2015,Babu_SoftMat23} and also in understanding the mechanism of shear jamming in dense suspensions~\cite{Mike2024,Santra_PRR2025}. In this work we have investigated the effects of polydispersity on the critical transition of cluster rigidity in terms of an order-parameter defined by the fraction of particles in rigid cluster, $f_\text{rig}$, same as that considered by \citet{Santra_PRR2025} in there study on bidisperse systems.
\begin{equation}
    f_\text{rig} = \langle\frac{1}{N}\sum\limits_{i=1}^N n_i\rangle
\end{equation}

\noindent here, $n_i=1$ if the particle is in a rigid cluster and $0$, otherwise, $\langle\cdot\cdot\rangle$ denotes ensemble average over several snapshots. While computing $f_\text{rig}$ we have discounted the particles which are in contact with at least one particle that is not a part of any rigid cluster, $i.e.$ the boundary particles of a rigid cluster are excluded from the calculation. Qualitatively, this definition of $f_\text{rig}$ produces the same variation with $\phi$ as that where the boundary particles are included in the calculation~\cite{Santra_PRR2025}.  A critical phenomenon is also associated with a divergence of the susceptibility (fluctuation of an order-parameter) $\chi_\text{rig}$ at the critical point~\cite{Malakis_2017,Toral_1987,Maggi2021,Dashti_SciAdv23}, where, 
\begin{equation}
\chi_\text{rig} = N \langle\frac{1}{N}\sum\limits_{i=1}^N (n_i - f_\text{rig})^2\rangle
\end{equation}
In the subsequent section we discuss the scaling behavior of $f_\text{rig}$ and $\chi_\text{rig}$ for the polydisperse systems and compared that with the bidisperse cases. 

\begin{figure*}[ptbh]
    \centerline{
    \begin{tabular}{c c}
        \includegraphics[width=87mm]{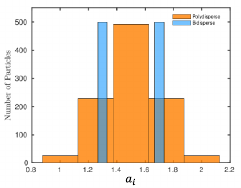} 
        & \includegraphics[width=90mm]{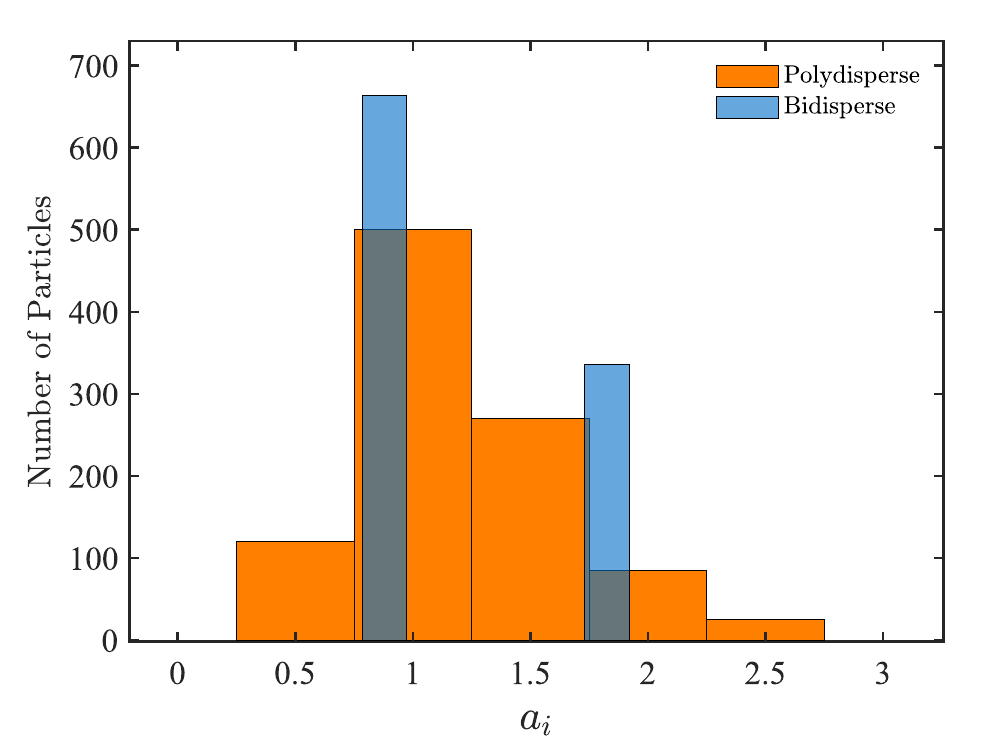} \\
        (a) & (b) \\
    \end{tabular}
    }
    
   \caption{\small{Histograms of polydisperse suspensions, following (a) normal distribution and (b) log-normal distribution, compared to the statistically equivalent bidisperse systems.} 
\label{fig:hist}}
\end{figure*}

\begin{figure*}[ptbh]
    \centerline{
    \begin{tabular}{c c}
        \includegraphics[width=85mm]{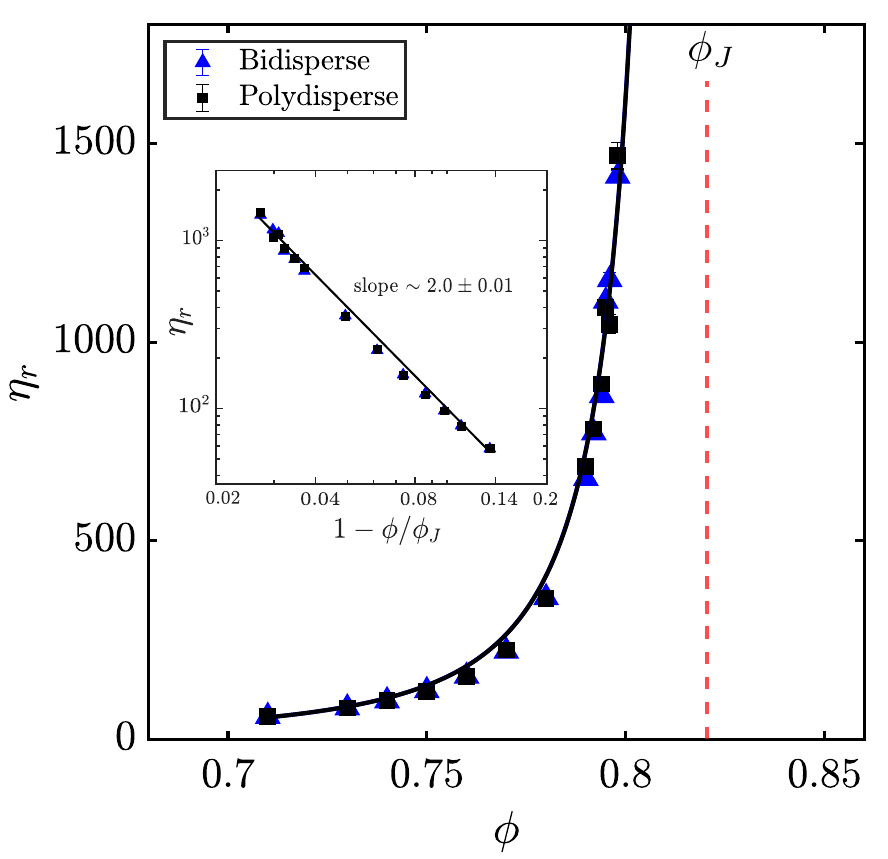} 
        & \includegraphics[width=85mm]{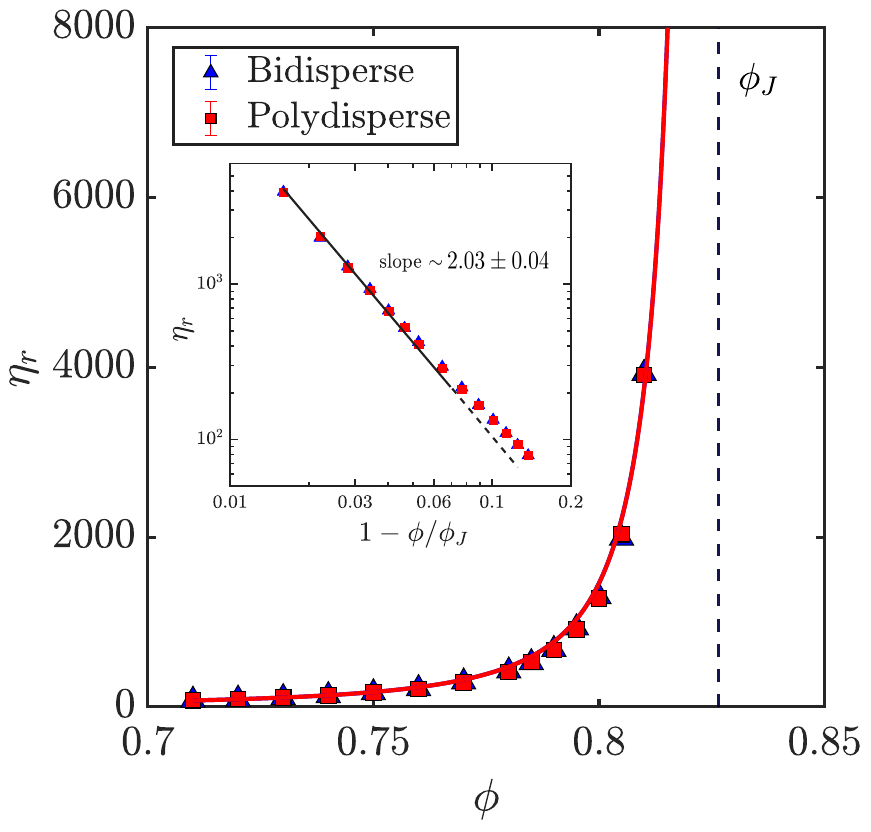} \\
        (a) & (b) \\
    \end{tabular}
    }
    
 \caption{\small{Divergence of relative viscosity $\eta_r$ as a function of packing fraction $\phi$ for (a) normal and its statistically equivalent bidisperse distributions with polydispersity index $\alpha=0.133$, at stress $\sigma=25$ and for (b) log-normal and its statistically equivalent bidisperse distributions, with $\alpha=0.374$, at $\sigma=100$. The lines represent M-P fit: $\eta_r \sim \left(1-\frac{\phi}{\phi_J(\sigma)}\right)^{-b}$, where $\phi_J(\sigma)$ is the packing fraction in the shear jamming limit and $b$ is the scaling exponent. {\color{black} The estimated values of $\phi_J$ in sub-figure (a) and (b) are $0.8221\pm 0.0017$ and $0.8231\pm 0.0012$, respectively. The M-P fit is illustrated in the inset by plotting $\eta_r$ as a function of $(1-\phi/\phi_J)$ on a log-log scale.}}  
\label{fig:etavsphi_polybidi}}
\end{figure*}

To validate the bulk rheological behavior of a 2D polydisperse suspension, we have compared their viscosity with that of a statistically equivalent bidisperse suspension. Here, two suspensions are defined to be statistically equivalent if $\langle a\rangle$, $s.d.$ and $s$ are identical for both the systems, as illustrated in Figs.~\ref{fig:hist} (a) and \ref{fig:hist}~(b). {\color{black}It is worth noting here that we have considered a truncated normal and log-normal distribution to keep the size of the smallest particle well above $1 \,\mu m$, so that the Brownian forces are still negligible.}  In Fig.~\ref{fig:etavsphi_polybidi} we show that the divergence of the relative viscosity ($\eta_r=\eta_\text{suspension}/\eta_0$) as a function of the packing fraction is identical for the polydisperse and statistically equivalent bidisperse suspensions in 2D at stresses $\sigma=25$ and $\sigma=100$, for both normal and log-normal distributions. This is in agreement with that observed in case of the suspensions simulated in 3D~\cite{Pednekar2018}. Furthermore, Fig.~\ref{fig:etavsphi_polybidi} (a) and \ref{fig:etavsphi_polybidi} (b) demonstrate that the viscosity divergence can be described by the Maron-Pierce (M-P) model ($\eta_r \sim \left(1-\frac{\phi}{\phi_J(\sigma)}\right)^{-b}$, where $\phi_J(\sigma)$ is the shear jamming fraction {\color{black}and $b$ is the scaling exponent obtained by fitting the simulation data}), similar to a 3D suspension~\cite{Nelya2023,Pednekar2018}. {\color{black} The values of $\phi_J$ and $b$ estimated for the data in Fig.~\ref{fig:etavsphi_polybidi} (a) and Fig.~\ref{fig:etavsphi_polybidi} (b) are $\phi_J = 0.8221\pm 0.0017$, $b=2.0\pm0.01$, and $\phi_J=0.8231\pm 0.0012$, $b=2.03\pm0.04$, respectively.  It is to be noted that while the M-P model fit the whole set of data for the normal distribution, in case of log-normal distribution the data far from the jamming point (\textit{i.e.} $\phi \ll \phi_J$) deviate from the M-P line, as shown in the insets of Fig.~\ref{fig:etavsphi_polybidi} (a) and (b).}

%%%%%%%%%%%%%%%%%%%%%%%%%%%%%%%%%%%%%%%%%%%%%%%%%%%%%%
\section{\label{sec:Results}Results and discussions}
%%%%%%%%%%%%%%%%%%%%%%%%%%%%%%%%%%%%%%%%%%%%%%%%%%%%%% 
\subsection{Scaling at the rigidity transition\label{sec:frig_scaling}}

\begin{figure*}[ptbh]
  \centerline{
 \resizebox{\textwidth}{!}{ \begin{tabular}{cc}
        \includegraphics[width=9.2cm,height=!]{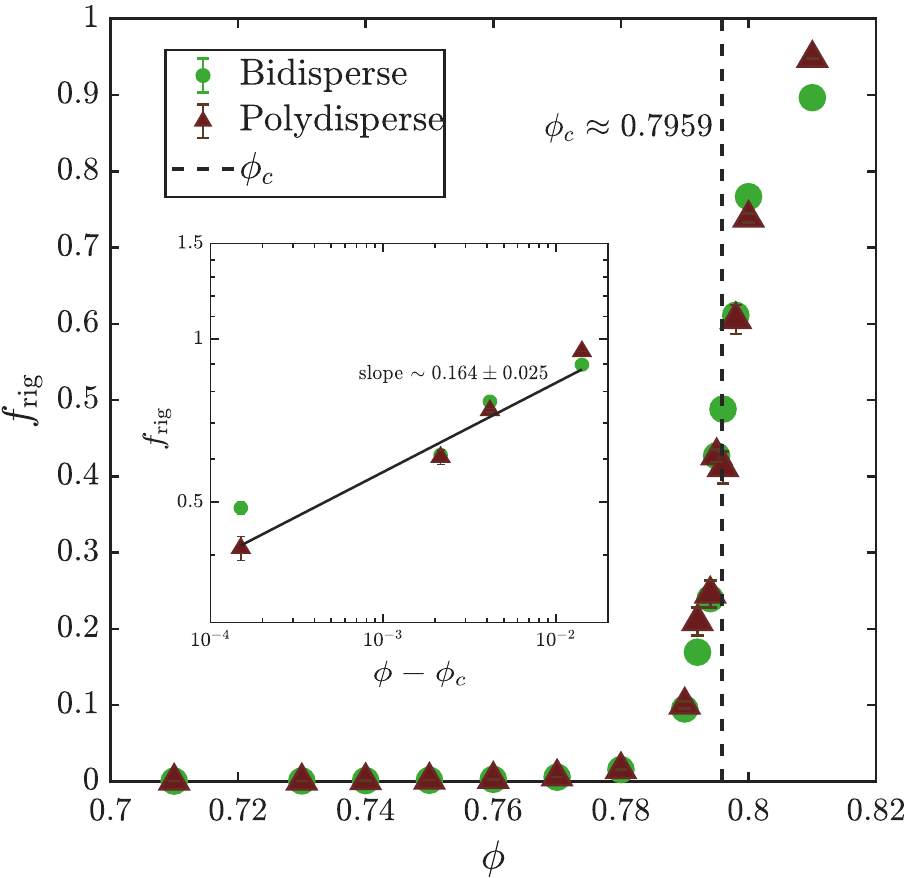} 
    & 
       \includegraphics[width=9cm,height=!]{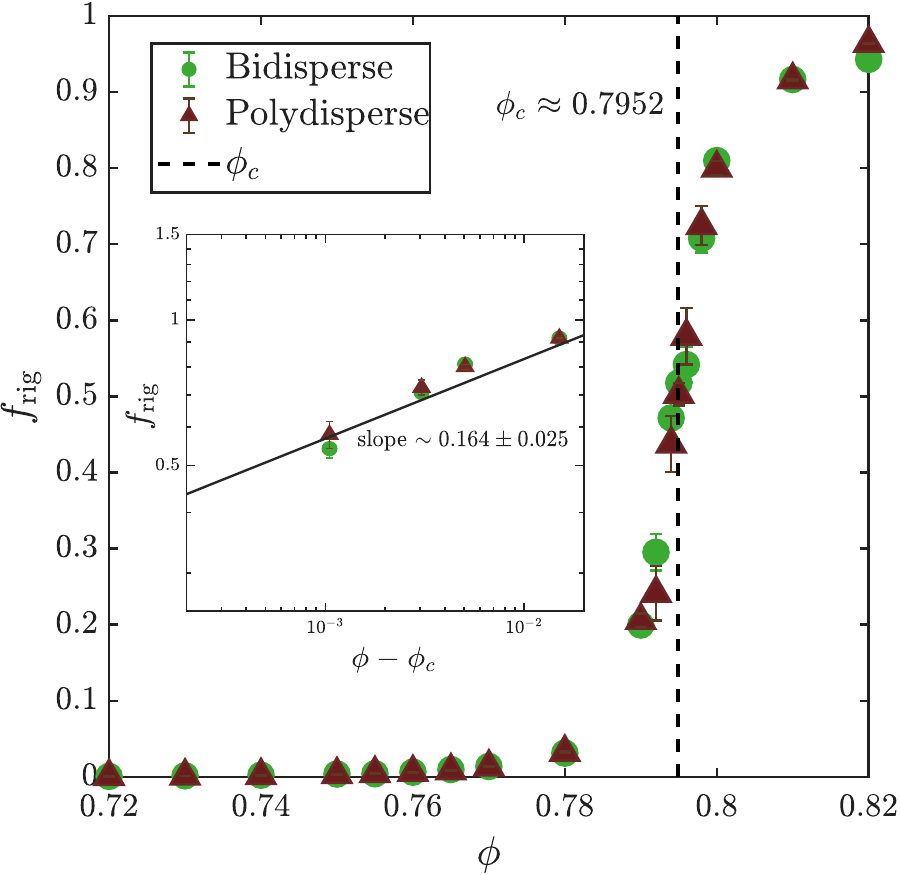} \\[5pt]
           (a)  & 
       (b) \\
       \multicolumn{2}{c}{\includegraphics[width=9cm,height=!]{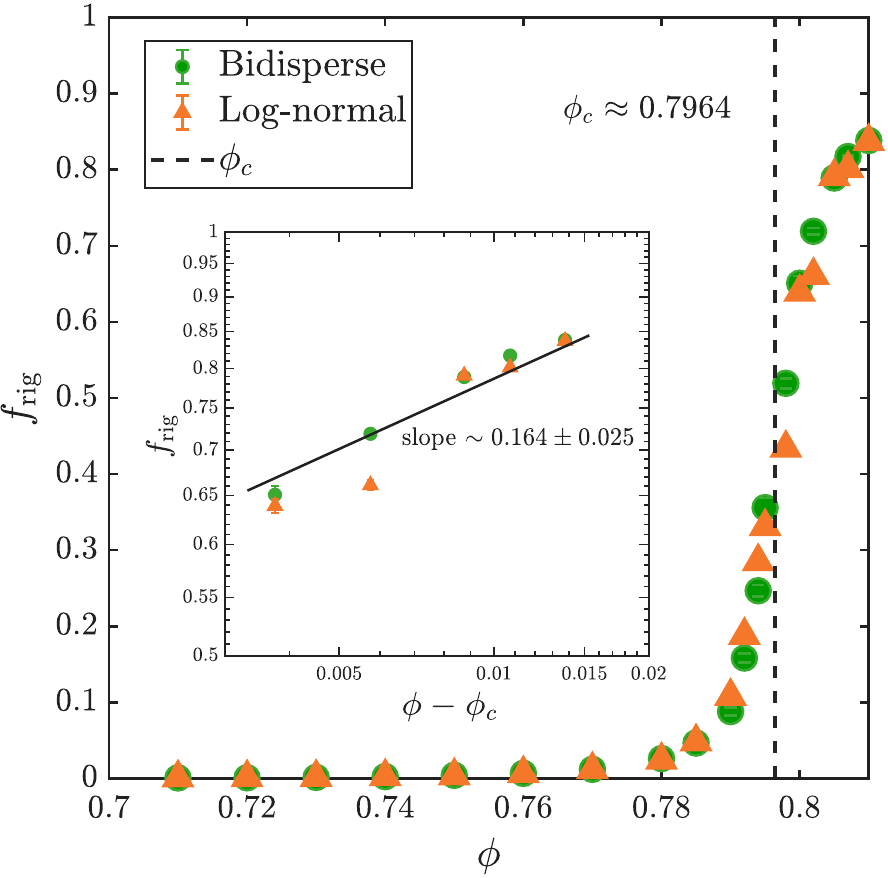} }\\[5pt]
       \multicolumn{2}{c}{(c)}     \\
  \end{tabular} }}
\caption{\small{Variation of the order-parameter $f_\text{rig}$ with solid packing fraction $\phi$ for normal and its statistically equivalent bidisperse distributions with $\alpha=0.133$, at stress (a) $\sigma=25$ and (b) $\sigma=100$. (c) Variation of  $f_\text{rig}$ with $\phi$ for log-normal and its statistically equivalent bidisperse distributions with $\alpha=0.374$, at $\sigma=100$. $\phi_c$ denotes the corresponding critical packing fraction. {\color{black}The insets show $f_\text{rig}$ as a function of $(\phi-\phi_c)$ on a log-log scale with the solid line representing a power-law fit with exponent $\beta=0.164\pm 0.025$.}} 
\label{fig:frig_bidipoly}}
\end{figure*}

\begin{figure*}[ptbh]
  \centerline{
 \resizebox{\textwidth}{!}{ \begin{tabular}{cc}
        \includegraphics[width=9.2cm,height=!]{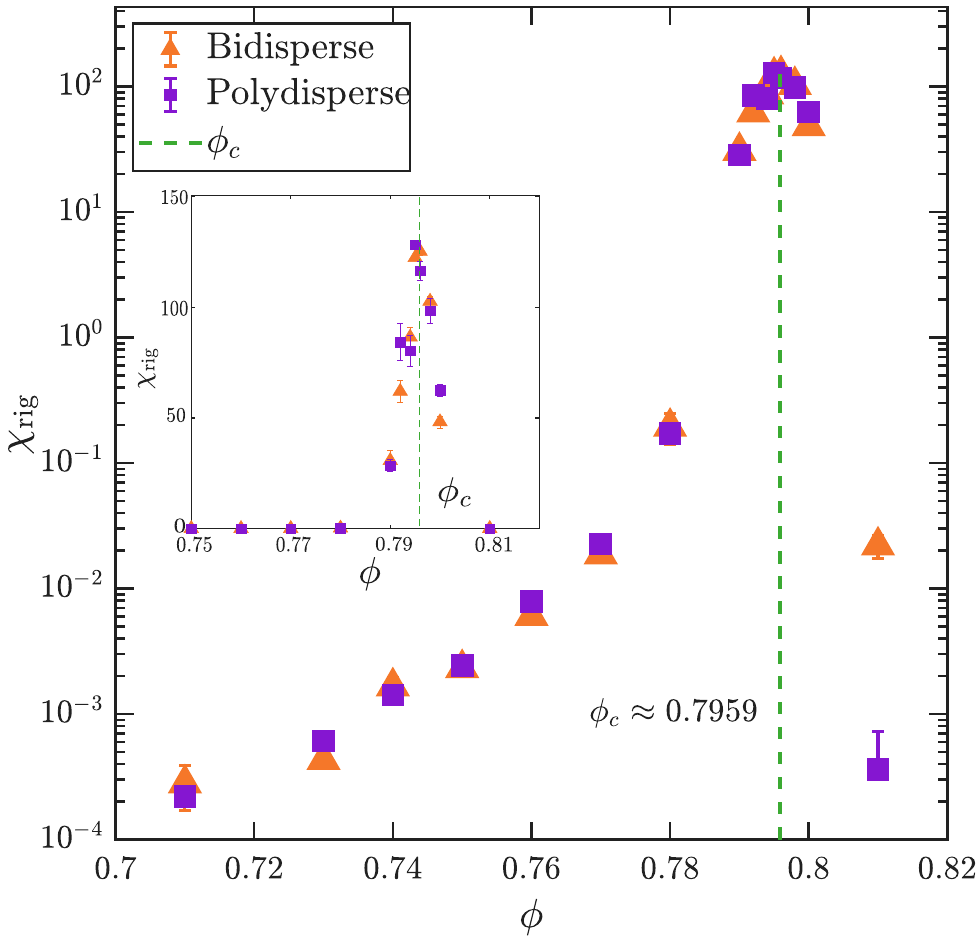} 
    & 
       \includegraphics[width=9.1cm,height=!]{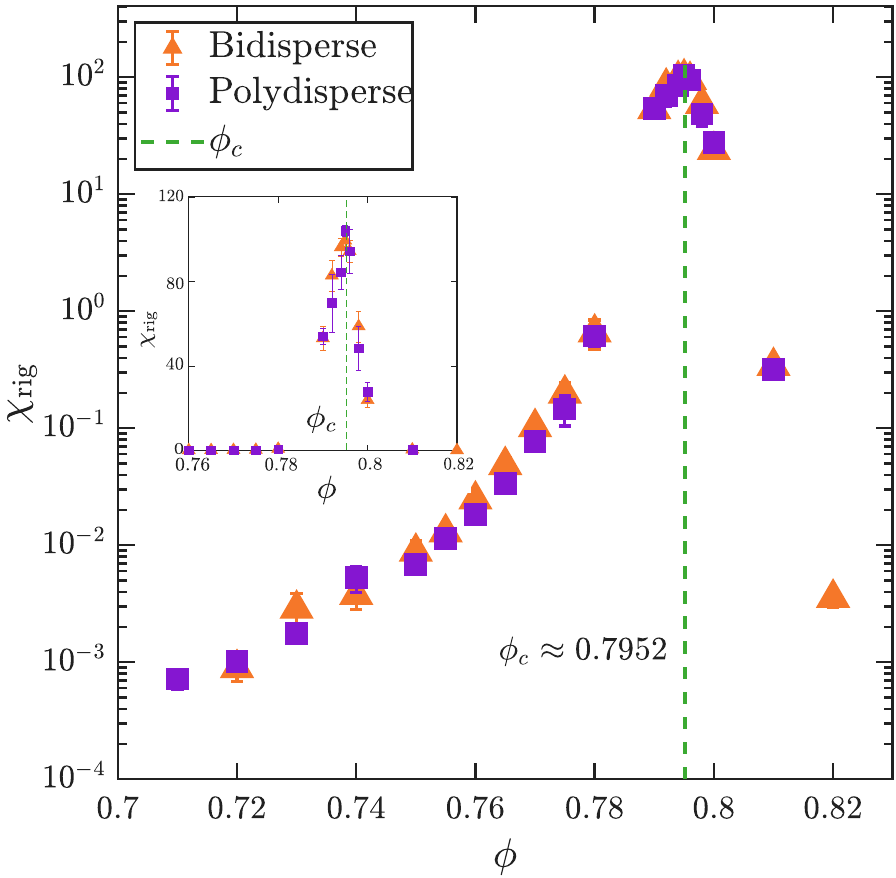} \\[5pt]
           (a)  & 
       (b) \\
       \multicolumn{2}{c}{\includegraphics[width=9cm,height=!]{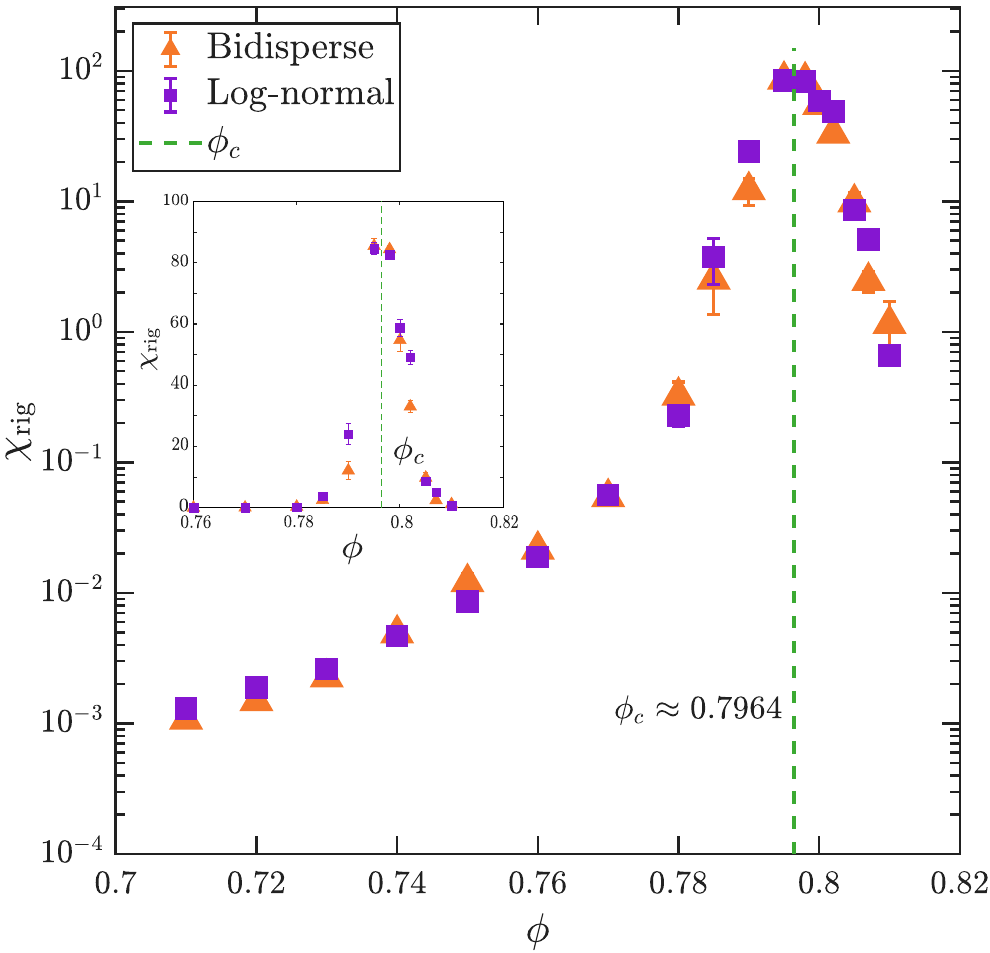} }\\[5pt]
       \multicolumn{2}{c}{(c)}     \\
  \end{tabular} }}
\caption{\small{Variation of the susceptibility $\chi_\text{rig}$ with solid packing fraction $\phi$ for normal and its statistically equivalent bidisperse distributions with $\alpha=0.133$, at stress (a) $\sigma=25$ and (b) $\sigma=100$. (c) Variation of  $\chi_\text{rig}$ with $\phi$ for log-normal and its statistically equivalent bidisperse distributions with $\alpha=0.374$, at $\sigma=100$. {\color{black}The vertical line at $\phi=\phi_c$ indicates the critical point at which $\chi_\text{rig}$ maximizes. The insets represent the corresponding data in linear scale.}} 
\label{fig:chirig_bidipoly}}
\end{figure*}

\begin{figure*}[ptbh]
    \centerline{
    \begin{tabular}{c c}
        \includegraphics[width=90mm]{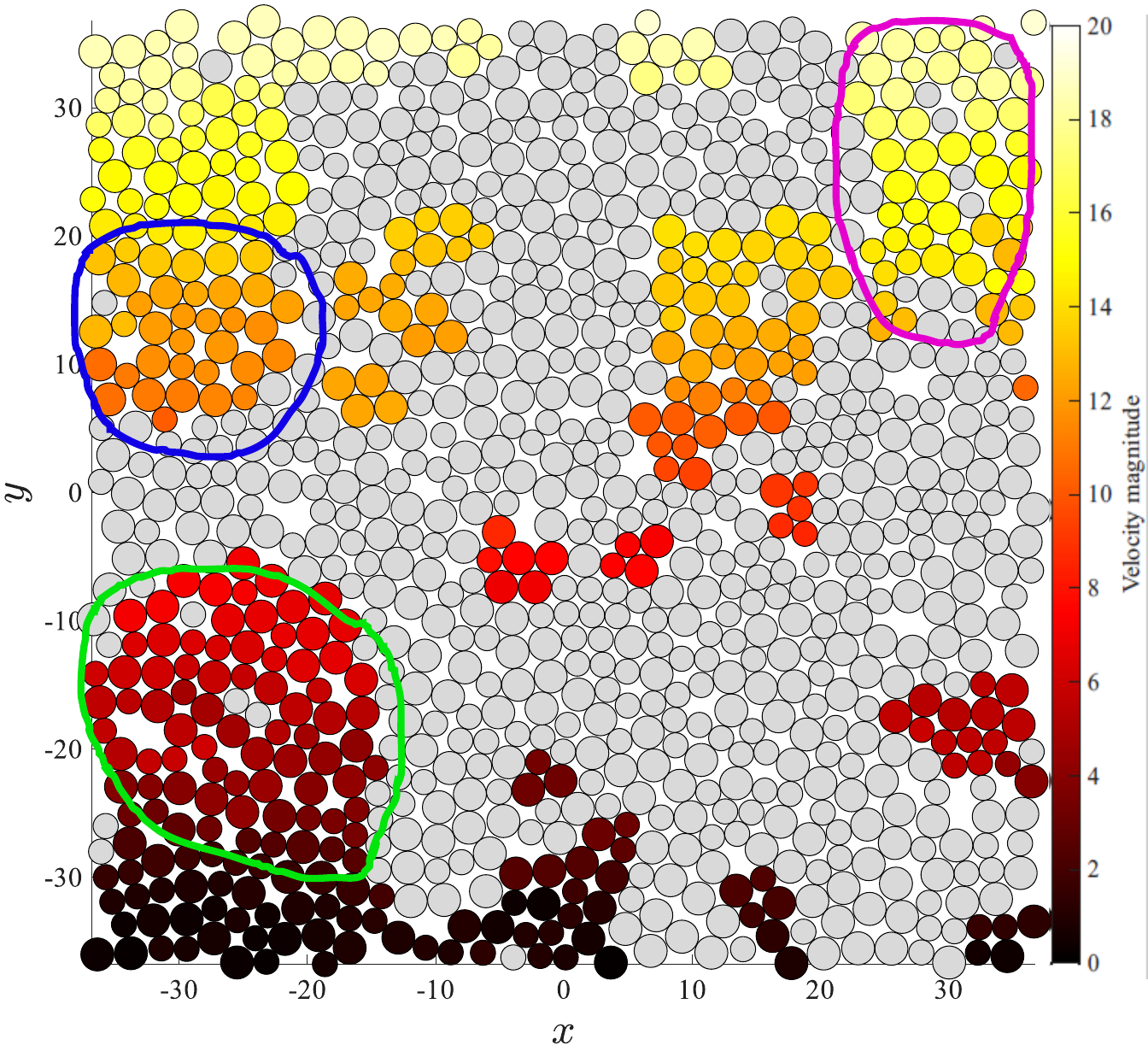} 
        & \includegraphics[width=90mm]{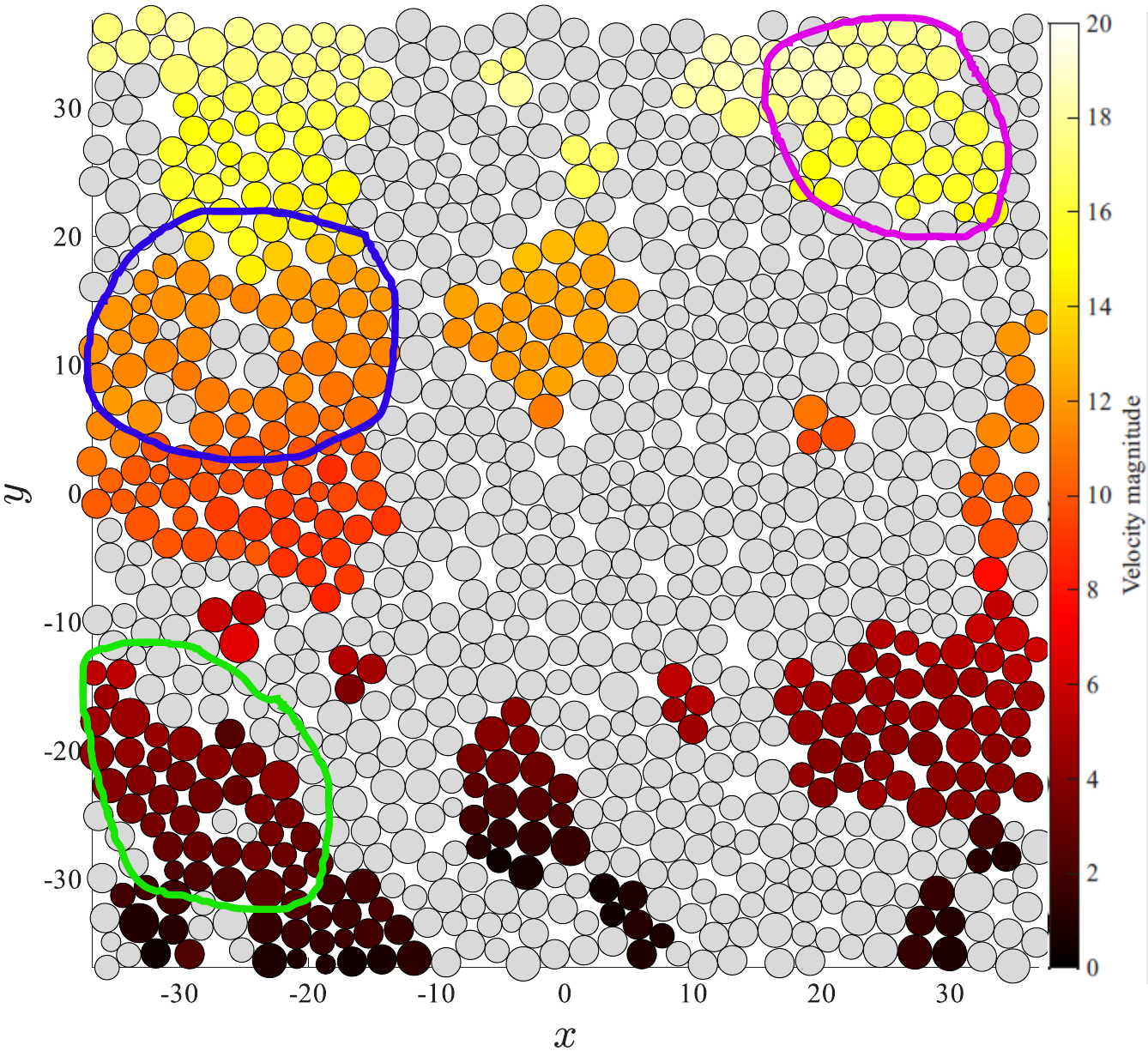} \\
        (a) & (b) \\
    \end{tabular}
    }
    
   \caption{\small{{\color{black}Snapshots of translational velocity distribution of the particles in rigid clusters shown by a colour map based on the magnitude of the particle velocity for (a) bidisperse suspension with $\alpha=0.133$, and (b) statistically equivalent polydisperse suspensions, at $\sigma=100$, $N=1000$. Blue, green and magenta circles are drawn to highlight similar locations in the simulation box for the two equivalent systems.}} 
\label{fig:veldist_bidi_poly}}
\end{figure*}

\begin{figure*}[ptbh]
  \centerline{
 \resizebox{\textwidth}{!}{ \begin{tabular}{cc}
        \includegraphics[width=9.5cm,height=!]{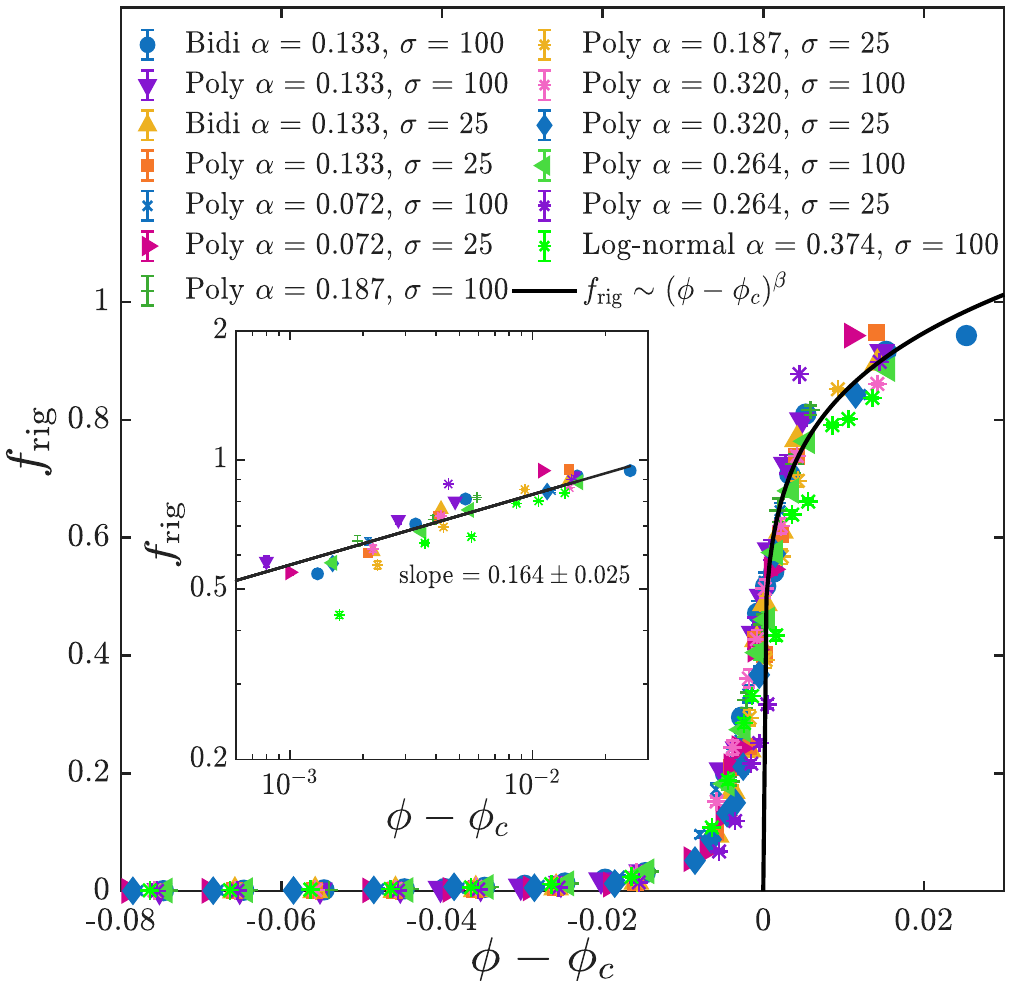} 
    & 
       \includegraphics[width=9.5cm,height=!]{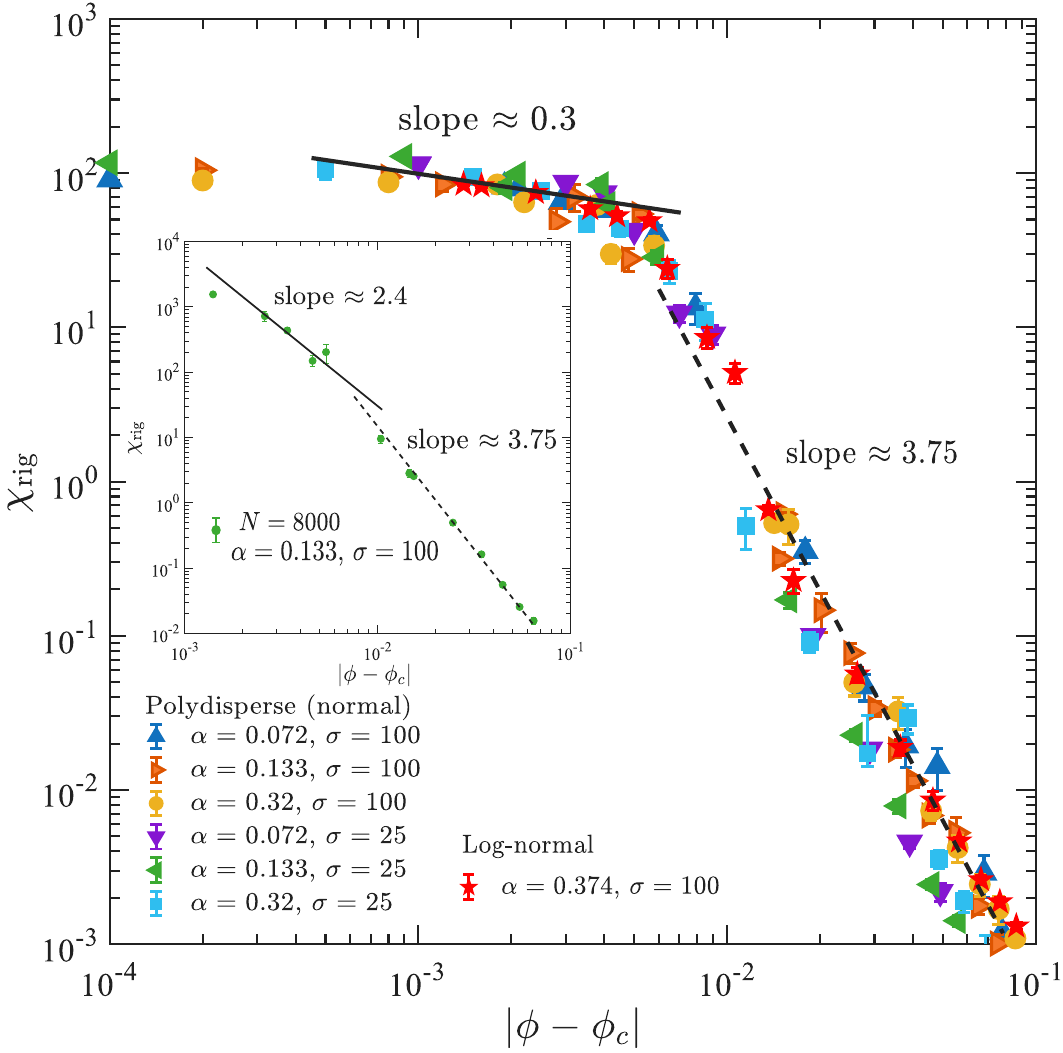} \\[5pt]
           (a)  & 
       (b) \\
  \end{tabular} }}
\caption{\small{Variation of (a) the order-parameter $f_\text{rig}$ and (b) susceptibility $\chi_\text{rig}$ as a function of the relative distance from the critical point ($\phi-\phi_c$) for different conditions of polydispersity at stress values of 100 and 25, for $N=1000$. {\color{black}(a) The inset presents $f_\text{rig}$ data on a log-log scale. The solid line represents a fit to the scaling relation $f_\text{rig}\sim (\phi-\phi_c)^{\beta}$, where $\beta=0.164\pm 0.025$. (b) The inset presents $\chi_\text{rig}$ data for $N=8000$ particles. The dashed line represents a power law with exponent $3.75$ for both the main plot and inset. The solid line in the main plot and inset represents a power law with exponent $\gamma\approx 0.3$ and $\gamma\approx 2.4$, respectively.}} 
\label{fig:frig_chirig_collapse}}
\end{figure*}

In this section, we characterize the growth of rigid clusters near shear jamming by investigating the scaling of the order-parameter $f_\text{rig}$ and susceptibility $\chi_\text{rig}$. Similar to bidisperse suspensions~\cite{Santra_PRR2025}, Figs.~\ref{fig:frig_bidipoly} show that $f_\text{rig}$ for polydisperse systems also increases with solid packing fraction $\phi$ and undergoes a sharp transition at $\phi=\phi_c$, where $\phi_c(<\phi_J)$ denotes the critical packing fraction. Notably, the variations of $f_\text{rig}$ with $\phi$ for the polydisperse suspensions are identical to the statistically equivalent bidisperse systems. Furthermore, the growth of $f_\text{rig}$ above $\phi_c$ can be well fitted by the scaling relation, $f_\text{rig}\sim[\phi-\phi_c]^{\beta}$, {\color{black} analogous to a critical transition where $f_\text{rig}$ is the order parameter, $\phi$ is the control parameter and and $\phi_c$ denotes the critical point~\cite{Sykes_1974,Stauffer1985,Pierre2006,Kohlhepp_PRB1992,Malakis_2017,Maggi2021}. The critical transition is further demonstrated in Fig.~\ref{fig:chirig_bidipoly} by the presence of a sharp peak in the variation of $\chi_\text{rig}$ as a function of $\phi$, for both polydisperse and bidisperse systems. The critical volume fraction $\phi_c$ is estimated by fitting a second order polynomial $g(x)$ to the $\log(\chi_\text{rig})$ versus $\phi$ data close to the peak, as indicated in Fig.~\ref{fig:chirig_bidipoly}, and subsequently equating $g'(x) = 0$ to evaluate the value of $x (=\phi_c)$ at which $\chi_\text{rig}$ becomes maximum. The uncertainty in the values of $\phi_c$ is determined from the estimated error in the fitting parameters and is found to be negligible (order of $\sim 10^{-5}$). The exponent $\beta$ is evaluated by a linear fit to the $f_\text{rig}$ data on a log-log scale as indicated in the insets of Fig.~\ref{fig:frig_bidipoly}. It is to be noted that we have evaluated $\beta$ by fitting $f_\text{rig}$ for all the different cases of polydisperse suspensions, taken together (this is further discussed in the context of Fig.~\ref{fig:frig_chirig_collapse} (a), in the subsequent paragraphs). Notably, the estimated value of $\beta$ is $0.164\pm 0.025$, which is in agreement with the value of critical exponent $\beta=5/36\approx 0.14$ for standard 2D percolation~\cite{Stauffer1985,Sykes_1974,Pierre2006}. Recently, \citet{Goyal_JOR2024} studied the percolation of 4-neighbor particles at DST in 3D systems and found the percolation probability (order parameter) to grow with an exponent of $\beta\approx 0.18$, which is far from the theoretical prediction~\cite{Stauffer1985,Scott1976}. On the contrary, considering $f_\text{rig}$ as the order parameter for a 2D system we show that there is a critical transition near shear jamming which could be well described by 2D percolation theory.} Similar to the order-parameter, the susceptibility variation is also identical for the polydisperse and the statistically equivalent bidisperse suspensions. Such identical variations of the order-parameter and susceptibility clearly indicate that not only the bulk rheology but also the growth and development of the flow microstructure in a shear thickening suspension of polydisperse particles can be effectively predicted from the microstructure of a relatively simple, equivalent bidisperse system. {\color{black}Equivalence in the maximum packing fraction and macroscopic rheological properties between different forms of polydisperse suspensions, mostly in the limit of frictionless jamming are already well explored~\cite{Desmond_2014,Chi2015}. However, such equivalence in the microstructural properties, such as cluster distribution or rigidity transitions were not so well reported. Previously, \citet{Pednekar2018} have shown that particles of a given size identically contributes to the hydrodynamic viscosity for two statistically equivalent systems, suggesting a similar stress environment for a particle of a given size in polydisperse and equivalent bidisperse suspensions. However, its consequence on the cluster rigidity and critical transition near shear jamming was not well understood. In the present study we have investigated these aspects in the framework of 2D percolation transition. Additionally, the simulation results suggest that at a coarse grained level the local stress becomes insensitive to the local particle composition. This is illustrated in Fig.~\ref{fig:veldist_bidi_poly}, where we have presented simulation snapshots of a bidisperse and statistically equivalent polydisperse system to show the comparison of the translational velocity distribution (which is related to the local stress distribution) in their respective rigid zones. As shown in Fig.~\ref{fig:veldist_bidi_poly}~(a) and \ref{fig:veldist_bidi_poly}~(b), for both the systems, magnitude of the velocity varies within the same range. Notably, at similar locations (highlighted by blue, green and magenta circles) within the simulation box, the velocity distributions of the particles for both the systems are nearly identical even though the local particle composition in the highlighted regions are very different. It is to be noted that since Fig.~\ref{fig:veldist_bidi_poly} displays a single simulation snapshot, exact match in the velocity distribution between the statistically equivalent systems is not observed, for which ensemble averaging over large number of snapshots is required.} The similarity between the microstructural features of a polydisperse suspension with that of its statistically equivalent bidisperse system is further demonstrated in the Appendix by the identical behavior of gradient direction velocity correlation functions.

Moreover, we can construct a master-curve, as shown in Fig.~\ref{fig:frig_chirig_collapse} (a), to collapse the values of the order-parameter over a range of packing fraction, at different conditions of polydispersity, by plotting $f_\text{rig}$ as a function of the relative distance from the critical point ($\phi-\phi_c$). {\color{black}The inset in Fig.~\ref{fig:frig_chirig_collapse}~(a) further demonstrates the data collapse on a log-log scale. As mentioned in the previous paragraph, the data set is fitted on a log-log scale to extract a value of the power law exponent $\beta$, which is found to be $0.164\pm 0.025$, consistent with the 2D percolation exponent. The susceptibility data in Fig.~\ref{fig:frig_chirig_collapse}~(b) are also found to collapse on a master-curve and show a crossover in the scaling exponent for the variation of $\chi_\text{rig}$ as a function of $|\phi-\phi_c|$. Far from the critical packing fraction ($|\phi-\phi_c|\gg 0$) $\chi_\text{rig}$ is found to scale with a power law exponent of $\gamma\approx 3.75$, as shown by the dotted line. Whereas, close to the critical point (\emph{i.e.} $|\phi-\phi_c| < 10^{-2}$) the scaling exponent ($\gamma$) is much lower, $\gamma\approx 0.3$, as shown in the main plot of Fig.~\ref{fig:frig_chirig_collapse}~(b) for $N=1000$. However, it is noteworthy that for the larger system size, \emph{i.e.} $N=8000$, the data show a crossover in the scaling exponent from $\gamma\approx 3.75$ to $\gamma\approx 2.4$, as presented in the inset of Fig.~\ref{fig:frig_chirig_collapse}~(b). Notably, $\gamma=2.4$ is in agreement with the critical exponent for susceptibility, $\gamma=43/18\approx 2.4$, in 2D percolation transition~\cite{JaeDong2011,Stauffer1985}. \citet{Santra_PRR2025} have carried out a similar analysis to show that the scaling exponent $\gamma$ is approximately equal to $ 7/4$ close to the critical transition, consistent with the 2D Ising model. However, they have extracted such a value for a relatively smaller system size. From the present analysis it is clear that with increasing system size the susceptibility exponent for the rigidity transition approaches towards a value consistent with the critical exponent in 2D percolation. Furthermore, we have carried out finite size scaling analysis to show that the 2D percolation exponents for the order-parameter and susceptiblity actually work very well by demonstrating a good data collapse for the scaling functions, which is discussed in the subsequent section.

Nevertheless, the key observation from the results presented in Fig.~\ref{fig:frig_chirig_collapse} (a) and \ref{fig:frig_chirig_collapse} (b) indicates that shear thickening and jamming behavior in non-Brownian suspensions remain unaffected by the polydispersity and shape of the particle size distribution. This has major implications on the understanding of the microscopic flow behavior of polydisperse suspensions since we could predict the shear thickening and shear jamming limits by appropriately rescaling the data.} 

\subsection{Finite size scaling\label{sec:finite_scaling}}

\begin{figure*}[ptbh]
  \centerline{
 \resizebox{\textwidth}{!}{ \begin{tabular}{cc}
        \includegraphics[width=8cm,height=!]{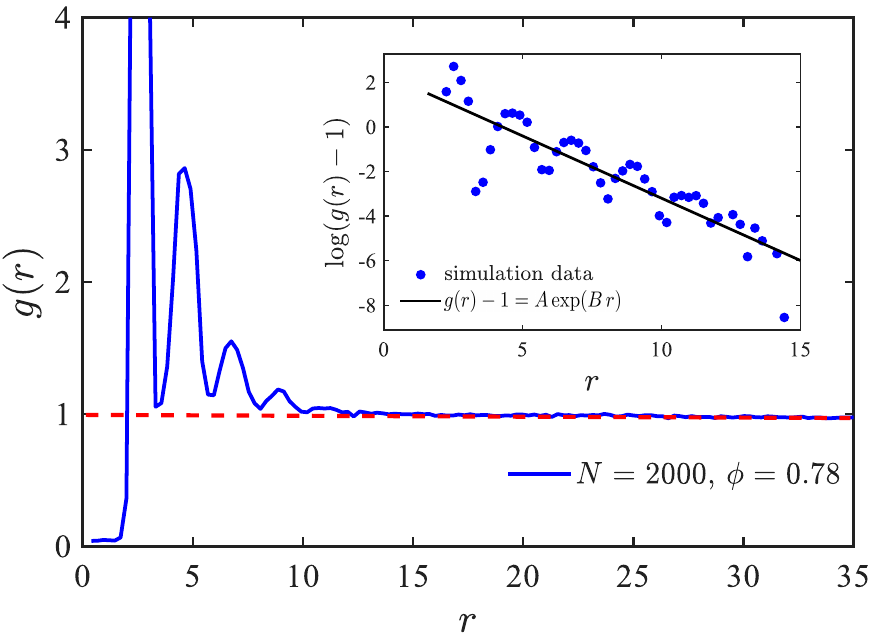} 
    & 
       \includegraphics[width=8.5cm,height=!]{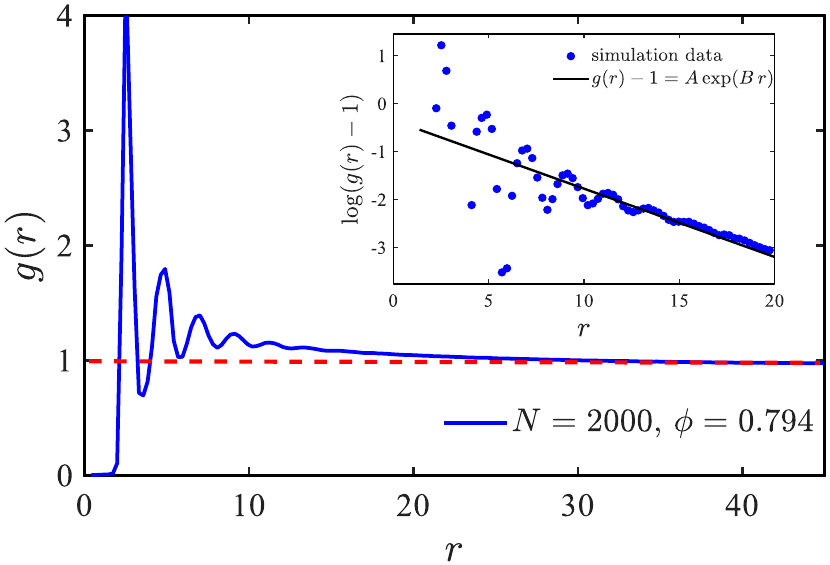} \\[5pt]
       \includegraphics[width=8cm,height=!]{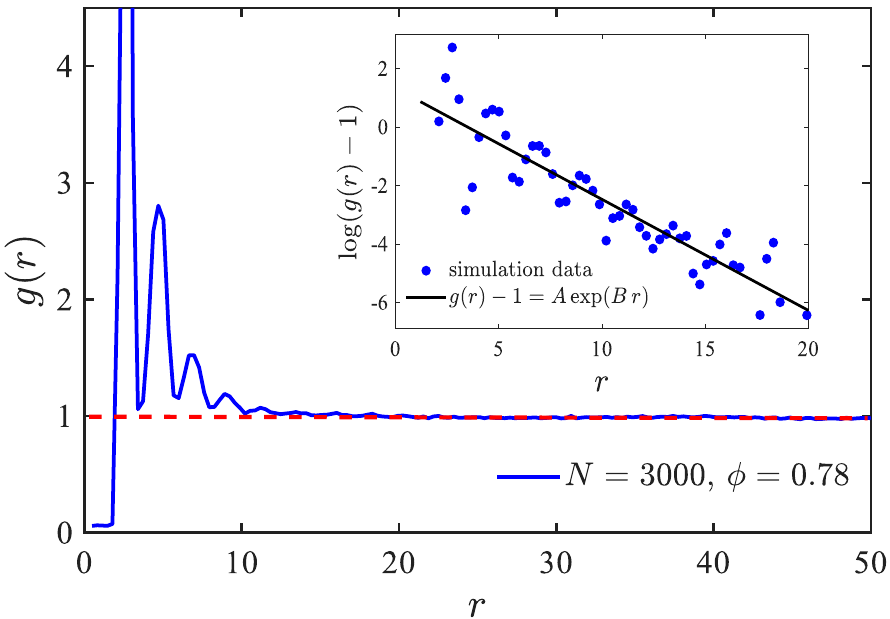} 
     &
       \includegraphics[width=8.4cm,height=!]{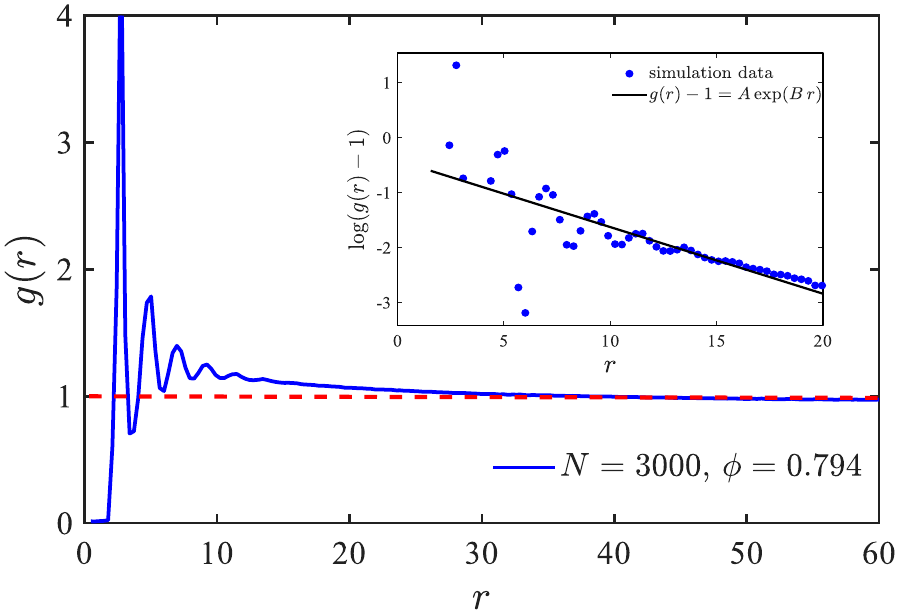} \\[5pt] 
      \includegraphics[width=8cm,height=!]{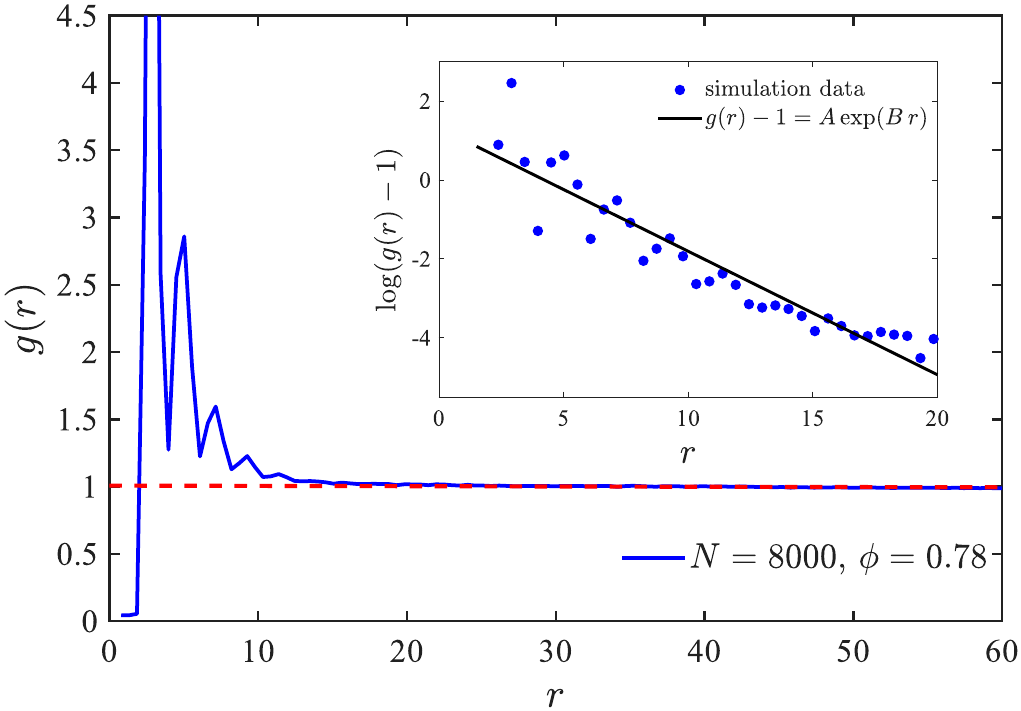} 
     &
       \includegraphics[width=8.4cm,height=!]{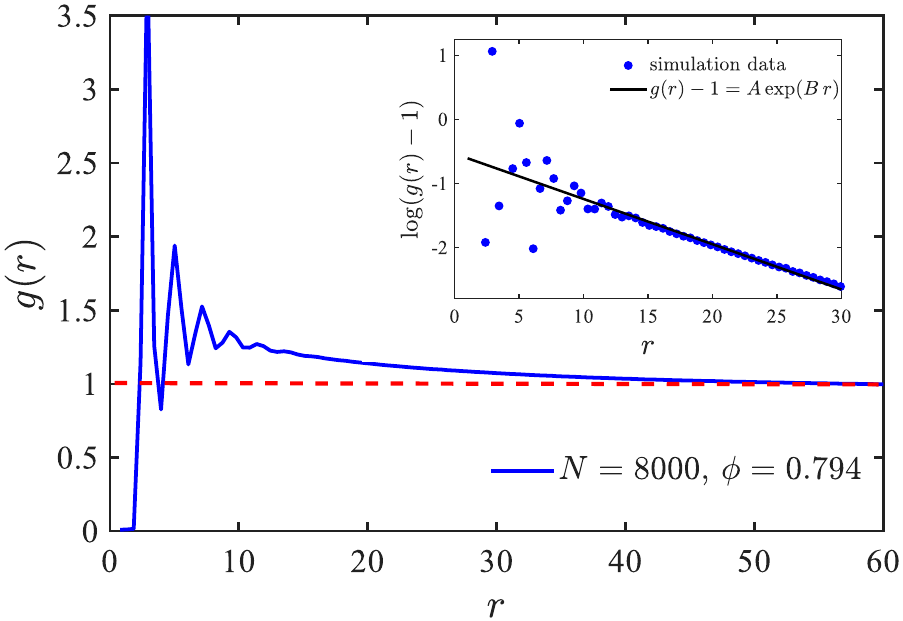} \\[5pt]   
           (a)  & 
       (b) \\
  \end{tabular} }}
\caption{\small{{\color{black}Pair-correlation function for rigid particles at (a) $\phi=0.78$ and (b) $0.794$ for normally distributed polydisperse systems ($\alpha=0.133$) with number of particles $N=2000$, $3000$ and $8000$. All the data are computed at stress $\sigma=100$. Insets show the variation of $g(r)-1$ as a function of $r$ on a semi-log scale. The solid line in the inset represents an exponential fit of the form $g(r)-1=A\exp(B\,r)$.}} 
\label{fig:gr_rigid}}
\end{figure*}

\begin{figure*}[ptbh]
  \centerline{
 \resizebox{\textwidth}{!}{ \begin{tabular}{cc}
        \includegraphics[width=12cm,height=!]{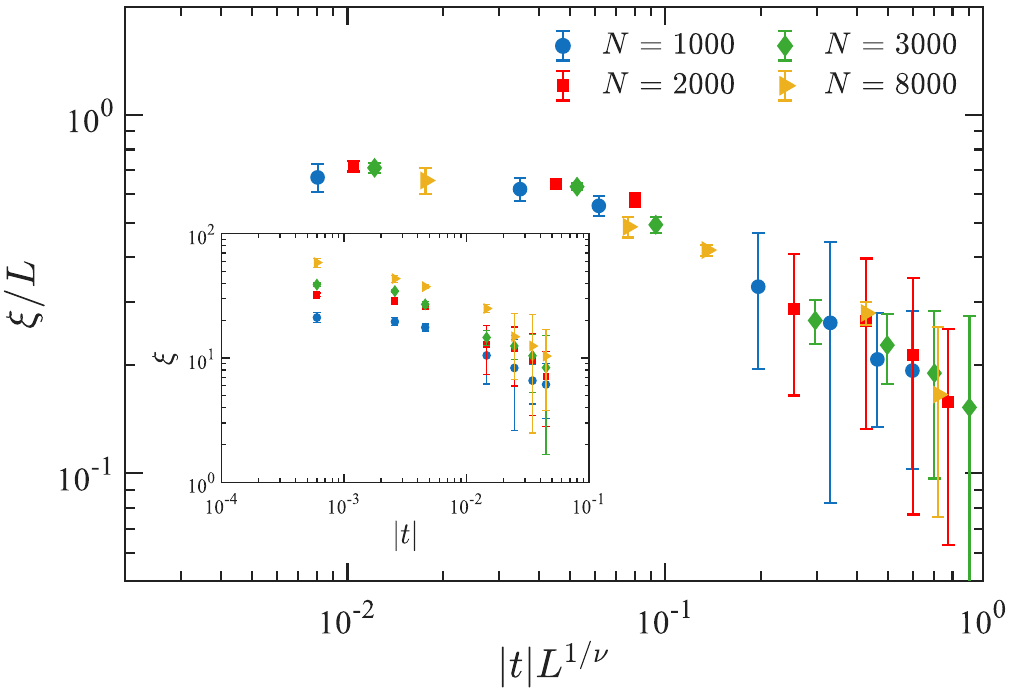} 
    & 
       \includegraphics[width=12cm,height=!]{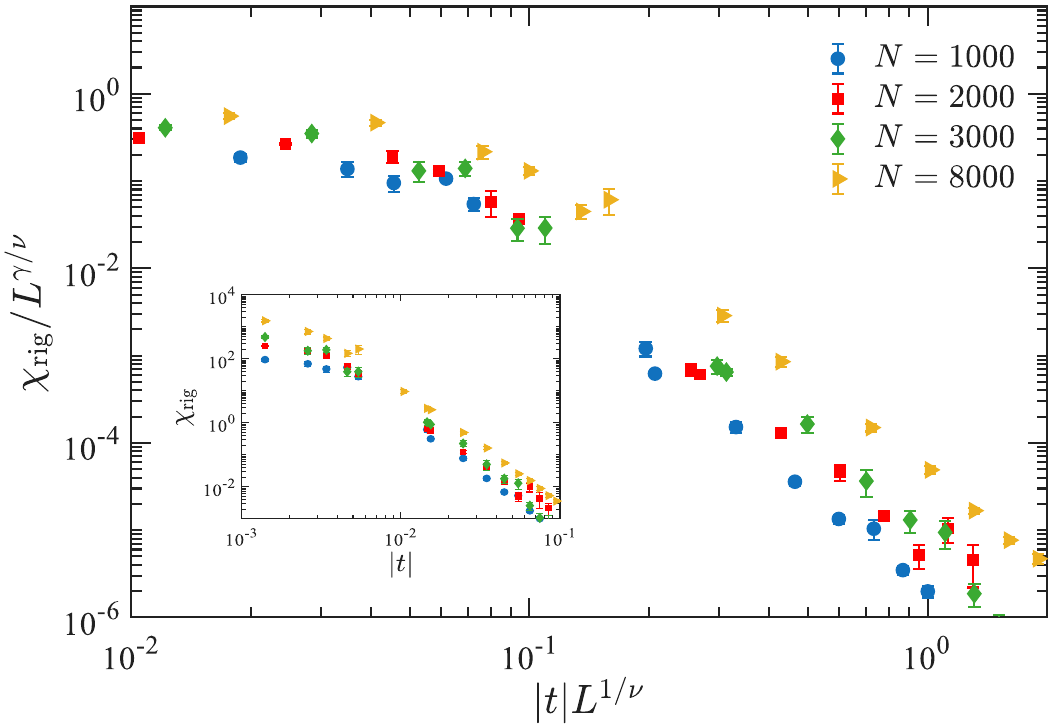} \\[5pt]
           (a)  & 
       (b) \\
       \multicolumn{2}{c}{\includegraphics[width=12cm,height=!]{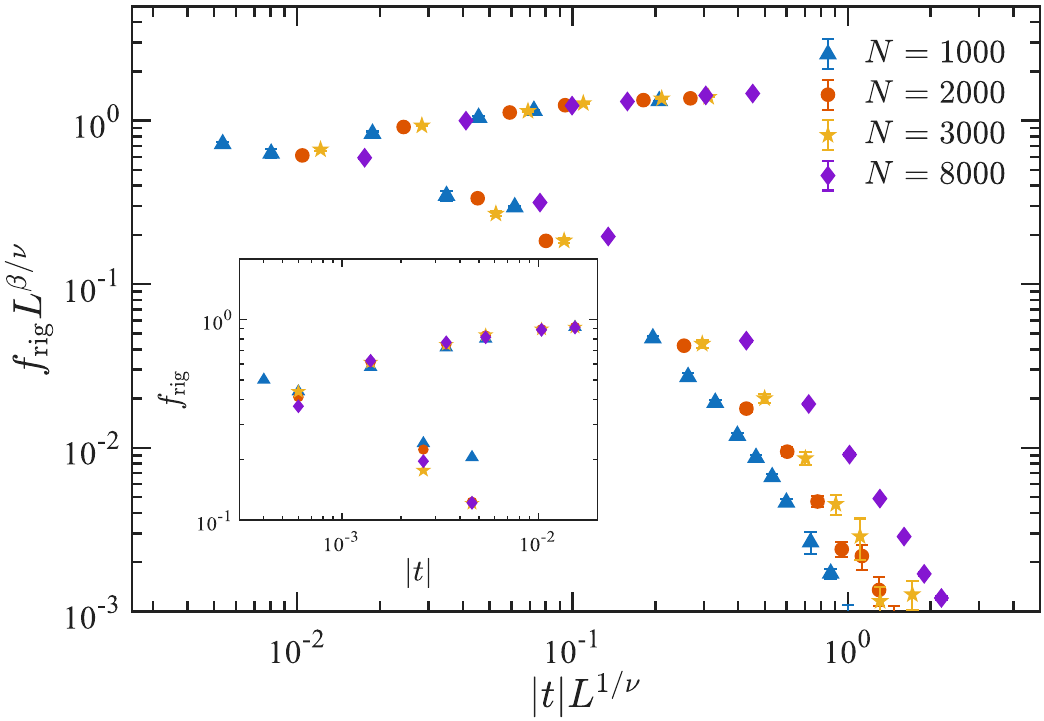} }\\[5pt]
       \multicolumn{2}{c}{(c)}     \\
  \end{tabular} }}
\caption{\small{{\color{black}Finite size scaling of the correlation length $\xi$, susceptibility $\chi_\text{rig}$, and order-parameter $f_\text{rig}$ for polydisperse suspensions with number of particles $N=1000$, $2000$, $3000$ and $8000$, $\alpha=0.133$ at stress $\sigma=100$. Collapse of data for different system sizes are observed with scaling exponents $\beta\approx 0.14$, $\gamma\approx 2.4$ and $\nu\approx 1.33$. The insets display the corresponding unscaled data.}}} 
\label{fig:finitesize}
\end{figure*}

Critical phase transitions in equilibrium systems are always associated with a diverging correlation length~\cite{Privman1984,Zhang2017,cardy1996scaling,Blume1971} which results into the universal scaling relations for the order-parameter and susceptibility as discussed in the previous section. While our systems of shear thickening suspensions are fundamentally out of equilibrium, the observation of a critical transition for such a system is in itself a noteworthy finding. {\color{black}Furthermore, the agreement with the scaling exponents of the 2D percolation transition suggests that the fraction of particles in rigid clusters ($f_\text{rig}$) can be used as an order-parameter analogous to the percolation probability (which is the typical order-parameter for percolation transition) to explain the non-equilibrium phase transition close to the shear jamming in dense suspensions, where the system undergoes a transition from a flowable to a non-flowable state.} While approaching a critical point, the correlation length ($\xi$), associated with the order parameter, is found to increase as $\xi \sim |t|^{-\nu}$, where $\nu$ is a critical exponent and $|t|$ is defined as the relative distance from the critical point in the limit of an infinite system size, \textit{i.e.} in the context of the present study $|t|=|\phi-\phi_{c_{\infty}}|$, where $\phi_{c_{\infty}}$ is the critical packing fraction for an infinitely large system~\cite{Privman1984,cardy1996scaling}. This implies that for a system with linear dimension $L$, the correlation length at the critical transition becomes $\xi = L$, as $L\rightarrow \infty$. For finite size systems, the divergence of the correlation length near the critical point is characterized by the following finite size scaling relations~\cite{Privman1984,Zhang2017,cardy1996scaling,Barmatz2007,Gasparini2008,Wilson1971},
\begin{eqnarray}
    \xi(t,L) &=& L \mathcal{G}(t\,L^{1/\nu}) \\
    f(t,L) &=& L^{-\beta/\nu} \mathcal{E}(t\,L^{1/\nu}) \\
    \chi(t,L) &=& L^{\gamma/\nu} \mathcal{F}(t\,L^{1/\nu}) 
\end{eqnarray}\label{eq:finite_size_scale}
{\color{black}Here, $\mathcal{G}$, $\mathcal{E}$ and $\mathcal{F}$ are the scaling functions for the correlation length, order-parameter and susceptibility, respectively, and $t\,L^{1/\nu}$ is the corresponding scaling variable.} This allows us to determine the critical exponents, $\beta$, $\gamma$ and $\nu$, from simulations of systems with finite domain and to identify the characteristics of the critical phenomena. Here, we have investigated finite size scaling behavior of polydisperse suspensions near the rigidity transition by simulating systems with different number of particles ($N = 1000$, $2000$, $3000$ and $8000$) at stress $\sigma=100$. In our 2D simulations, since $N\propto L^2$, the system size is defined by $\sqrt{N}$. The correlation length for the rigid clusters can be estimated from a length scale associated with the cluster size. We have evaluated a length scale from the decay of the pair-correlation function, $g(r)$, for the rigid particles (identified by the pebble game algorithm) and used it as an estimate of the correlation length $\xi$ at different system sizes. This is illustrated in Figs.~\ref{fig:gr_rigid}, where the variations of $g(r)$ are shown for $\phi=0.78$ and $\phi=0.794$, at different system sizes. It is to be noted that for any given value of $\phi$, the pair-correlation function is normalized by the corresponding number density of rigid particles only (\emph{i.e.} non-rigid particles are excluded from the calculation). {\color{black}The long-range correlation length scale is extracted from the decay of the pair-correlation function by a method to identify the value of $r$ (radial distance) at which $g(r)$ settles to 1, which corresponds to the average number density of rigid particles. In this approach we have fitted the tail of the decay of $g(r)-1$ by an exponential function of the form $y=A\exp(B\,r)$, and estimated the value of $r$ at which $g(r)-1$ becomes $0.01$ (\emph{i.e.} when the correlation function has decayed by $99$\%) and considered that as an estimate of the correlation length ($\xi$). Uncertainty on the extracted value of $\xi$ is evaluated from the error estimate of the fitting parameters $A$ and $B$ by propagation of error. Insets in Fig.~\ref{fig:gr_rigid} display the variation of $g(r)-1$ as a function of $r$ on a semi-log scale, which has been used to fit the exponential function as discussed above. While this approach works very well for the larger system size, close to the critical transition (\emph{i.e.} $(\phi_{c_\infty}-\phi)< 10^{-2}$, for $\phi<\phi_{c_{\infty}}$, where $\phi_{c_{\infty}} \approx 0.7946$ is evaluated by extrapolation of the $\phi_c$ values for $N\rightarrow \infty$), the exponential fit may produce large errorbars for the fitting parameters in case of smaller systems  and far from $\phi_{c_{\infty}}$, where there are more fluctuations in $g(r)$. However, for the finite size scaling analysis it is necessary to estimate the correlation length close to the critical packing fraction, which could be estimated with negligible errorbars as presented in Fig.~\ref{fig:finitesize}~(a).}  Notably, the correlation length ($\xi$) extracted in this manner is found to increase with increasing system size and packing fraction. This suggests that there is a divergence in the correlation length as the packing fraction approaches the critical point ($\phi_{c_\infty}$) from below and at $\phi=\phi_{c_\infty}$, the correlation length $\xi\rightarrow\infty$ in the limit of infinite system size.  

{\color{black}The finite size scaling relations in Eqs.~(15)-(17) are tested for the normally-distributed polydisperse suspensions (with $\alpha=0.133$) by computing $\xi$, $f_\text{rig}$ and $\chi_\text{rig}$ for systems with $N=1000,\,2000,\, 3000$ and $8000$ particles. Fig.~\ref{fig:finitesize} (a) presents the scaling of $\xi/L$ as a function of the scaling variable $|t|\,L^{1/\nu}$, where $t=(\phi-\phi_{c_{\infty}})$ and $\phi_{c_{\infty}}\approx 0.7946$. A good data collapse is observed for $\xi/L$ by choosing a value of the scaling exponent $\nu\approx 1.33$. It should be noted that the unscaled data are well separated, as presented in the inset of Fig.~\ref{fig:finitesize}~(a), validating the choice of the scaling variables and parameters for obtaining the data collapse. As shown in Fig.~\ref{fig:finitesize} (b), the scaling of $\chi_\text{rig}/L^{\gamma/\nu}$ also indicates a data collapse for the different system sizes close to the critical transition. Notably, for $\chi_\text{rig}$ the quality of data collapse near the critical point (\textit{i.e.} $|t|\rightarrow 0$) improves with increasing system size. The corresponding values of the scaling exponents are chosen to be $\gamma\approx 2.4$ and $\nu\approx 1.33$, which are consistent with the susceptibility scaling exponent found in the cross-over regime near the critical point for $N=8000$ particles, as shown in the inset of Fig.~\ref{fig:frig_chirig_collapse}~(b), and finite size scaling of correlation length in Fig.~\ref{fig:finitesize} (a). It is noteworthy that $\nu=4/3\approx1.33$ is in agreement with the value of the critical exponent in 2D percolation transition~\cite{Stauffer1985,JaeDong2011}. Similarly, the finite size scaling for the order-parameter $f_\text{rig}$ is demonstrated in Fig.~\ref{fig:finitesize}~(c), where a collapse of data is observed near the critical point with $\beta\approx 0.14$ and $\nu\approx 1.33$. It is to be noted that there are two branches of the scaling function $f_\text{rig}L^{\beta/\nu}$, where the upper and the lower branch represents the data for $\phi>\phi_{c_\infty}$ and $\phi<\phi_{c_\infty}$, respectively. The inset in Fig.~\ref{fig:finitesize}~(b) and \ref{fig:finitesize}~(c) presents the corresponding unscaled data for $\chi_\text{rig}$ and $f_\text{rig}$, respectively. Moreover, similar to the scaling of the order-parameter and susceptibility discussed in section~\ref{sec:frig_scaling}, the consistency in the values of exponent $\beta$, $\gamma$ and $\nu$ with the exponents of 2D percolation transition suggests that the rigidity transition in shear thickening suspensions can be correlated to  a percolation transition~\cite{Mike2024,Goyal_JOR2024}.}

\subsection{Effects of polydispersity index and particle stiffness on $\phi_c$\label{sec:stiffness_effect}}

\begin{figure*}[ptbh]
  \centerline{
 \resizebox{\textwidth}{!}{ \begin{tabular}{cc}
        \includegraphics[width=9.2cm,height=!]{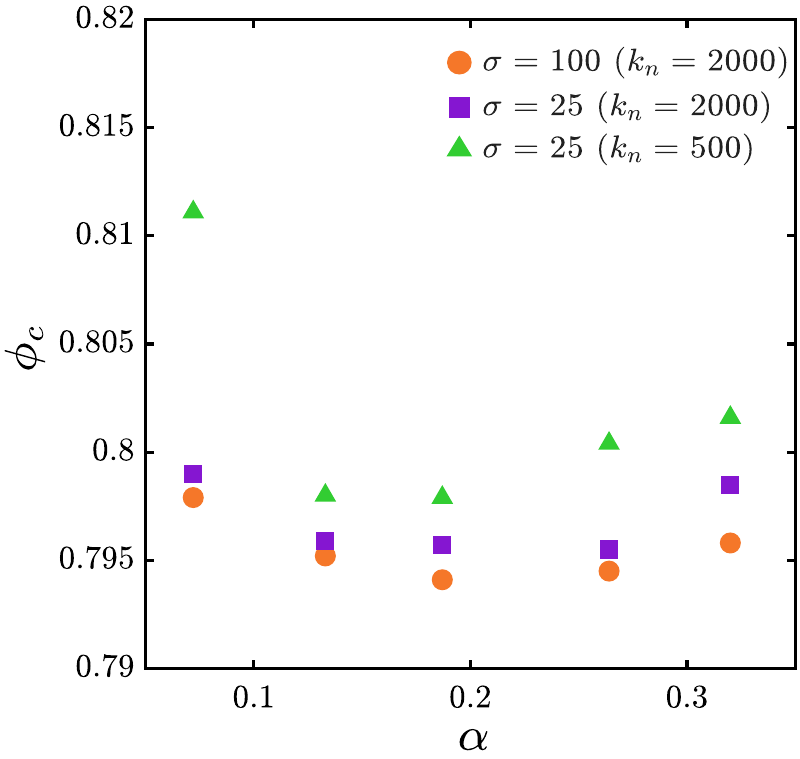} 
    &
        \includegraphics[width=9.2cm,height=!]{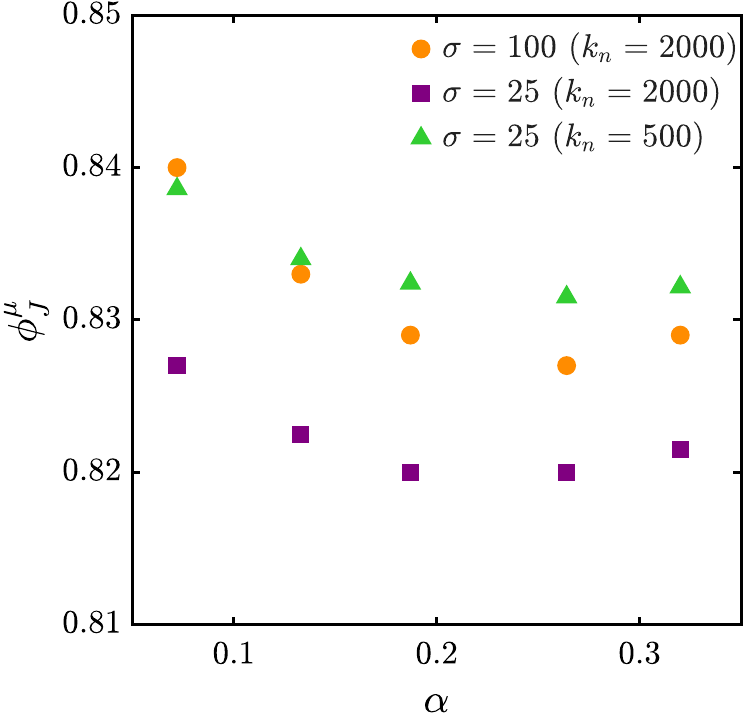}  \\[5pt]
         (a)  & 
       (b) \\
 \end{tabular} }}   
\caption{\small{Variation of (a) critical packing fraction ($\phi_c$) and (b) shear jamming fraction ($\phi_J^{\mu}$) as a function of polydispersity index $\alpha$ at different values of spring stiffness $k_n$.} 
\label{fig:stiffness_effect}}
\end{figure*}

According to the present model, the frictional contacts between the particles depend on the spring constants for the normal and tangential components of the contact force. While in this study we have considered the particles to be non-deformable rigid spheres, the stiffness of the particles may still be correlated to the stiffness of the springs at the frictional contact. This section discusses the effects of polydispersity index and spring constants on the critical packing fraction and shear jamming. As presented in Fig.~\ref{fig:stiffness_effect} (a), for a given value of stress ($\sigma$) and spring constant ($k_n$) the critical packing fraction $\phi_c$ varies non-monotonically with polydispersity index $\alpha$ and goes through a minima. The same trend is observed for the shear jamming fraction $\phi_J^{\mu}$, as shown in Fig.~\ref{fig:stiffness_effect} (b). {\color{black}This is in contrast to the nature of variation of the jamming threshold ($\phi_m$) observed for frictionless polydisperse systems~\cite{Desmond_2014,Chi2015}, where $\phi_m$ is found to monotonically increase with polydispersity index ($\alpha$) for positive skewness ($s>0$) and show only week dependence on $\alpha$ for $s<0$}. Notably, in the present system, while for the same value of $k_n$($=2000$) $\phi_c$ is smaller for the larger stress ($\sigma=100$) at all polydispersity indices (as shown in Fig.~\ref{fig:stiffness_effect} (a)), $\phi_J^{\mu}$ shows a opposite behavior where it is higher for $\sigma=100$ as compared to $\sigma=25$. It is noteworthy that by proportionally lowering the value of $k_n$ to 500 for $\sigma=25$, both $\phi_J^{\mu}$ and $\phi_c$ are found to be higher for $\sigma=25$ as compared to $\sigma=100$ at all values of $\alpha$ (except for very small $\alpha$). Thus with decreasing spring stiffness, both $\phi_J^{\mu}$ and $\phi_c$ increases at any given stress suggesting the suspensions to become more flowable with softer particle contacts. This can be correlated to increasing flowability of softer deformable particles which could easily squeeze past each other under shear flow as compared to more rigid particles. Moreover, the similar variation of $\phi_J^{\mu}$ and $\phi_c$ with $\alpha$ at large stresses suggests that the critical rigidity transition in shear thickening suspensions is simply a pre-cursor to the shear jamming transition.               

%%%%%%%%%%%%%%%%%%%%%%%%%%%%%%%%%%%%%%%%%%%%%%%%%%%%%%%%%%%%%%%%%%%%%%%%%%%%%%%%%%%%%
\section{\label{sec:Conclusions}Conclusions}
%%%%%%%%%%%%%%%%%%%%%%%%%%%%%%%%%%%%%%%%%%%%%%%%%%%%%%%%%%%%%%%%%%%%%%%%%%%%%%%%%%%%%
A discrete element method (LF-DEM) based numerical simulations are carried out for systems of highly concentrated polydisperse shear thickening suspensions in 2D to understand the effects of particle size distribution on rigidity transition near shear jamming in the limit of large shear stress. The relative viscosity of different polydisperse suspensions is found to be identical to that of their statistically equivalent bidisperse systems. Furthermore, the viscosity divergence in 2D polydisperse and statistically equivalent bidisperse suspensions follow the Maron-Pierce model, similar to 3D systems. In the limit of large stress values, shear jamming phenomenon in polydisperse suspensions is found to be preceded by a critical rigidity transition similar to that observed by \citet{Santra_PRR2025}, in their study on bidisperse suspensions. Pebble game algorithm is implemented to evaluate an order-parameter from the fraction of rigid particles, $f_\text{rig}$, and its fluctuation, $\chi_\text{rig}$, which show identical variation with solid packing fraction for the polydisperse and statistically equivalent bidisperse suspensions. Such characteristics illustrate the similarity in the micro-structure of polydisperse and bidisperse suspensions. This further implies that the evolution of the flow micro-structure in shear thickening suspensions of polydisperse particles can be understood from the properties of relatively simple bidisperse systems. {\color{black}One of the key findings from the present study suggests that the polydispersity only shifts the critical packing fraction $\phi_c$ without qualitatively changing the critical phenomenon, which could have been affected by the way particles pack with variation in polydispersity. However, within the limited scope of the present study with finite size systems we have not explicitly considered the effects of the nature of particle packing with polydispersity. To bring out any explicit dependence of the particle packing on the overall characteristics of the critical phenomenon, large scale simulations may be needed with larger system sizes. We may speculate that the nature of particle packing may appear as an external field variable in the scaling relations, similar to the role of a static friction coefficient investigated in another study~\cite{Santra_PRR2025}, but it warrants detailed investigation.} 

The critical transition observed in the polydisperse suspensions are characterized by determining the scaling behavior of $f_\text{rig}$ and $\chi_\text{rig}$. {\color{black}The values of the scaling exponents $\beta= 0.164\pm0.025$ and $\gamma\approx 2.4$, for $f_\text{rig}$ and $\chi_\text{rig}$, respectively, are found to be consistent with the 2D percolation transition. Furthermore, the values of the critical exponents $\beta$, $\gamma$ and $\nu$ are also approximated from the finite size scaling analysis of the correlation length, order-parameter and susceptibility.} In this context, we have used a protocol to compute the correlation length of the rigid clusters from the pair-correlation function of the rigid particles. {\color{black}Notably, by choosing the values of $\beta\approx 0.14$, $\gamma\approx2.4$ and $\nu\approx1.33$, which are consistent with the values of critical exponents for standard 2D percolation transition, we have shown good data collapse within the framework of finite size scaling ansatz. This essentially validates that the rigidity transition in shear thickening suspensions also corresponds to a percolation transition, which was qualitatively indicated in a previous study by \citet{Mike2024} for 2D suspensions, and quantitatively analyzed in an independent study by \citet{Goyal_JOR2024} for 3D systems by computing percolation probability of 4-neighbor particles.} Our investigation also indicates that the critical packing fraction $\phi_c$ is affected by the stiffness of the particles, moving the critical transition to higher packing fraction with decreasing spring constants. While only high stress values are considered in the present work to focus on the jamming transition well above the discontinuous shear thickening (DST) regime, flow behavior and rigidity transition for polydisperse suspensions at lower stresses would be even more interesting because of the effect of CST-DST transition, which would be addressed in the future work. Nevertheless, the present study would be the basis to experimentally explore critical transition in shear thickening suspensions of more complex systems with heterogeneous particle size and shapes. {\color{black}This is particularly important from the practical point of view since the ability to predict the critical volume fraction could prevent getting into undesirable jammed states in various industrial operations involving dense suspensions.} Moreover, we show that both the bulk rheology and microscopic statistical features of polydisperse suspensions could be predicted from the behavior of statistically equivalent bidisperse suspensions. We also demonstrate that rigidity transition in shear thickening suspensions with different particle size distributions has features which could be explained by a continuous phase transition model, such as, 2D percolation theory.

\linespread{1}\selectfont

\section*{Author Contributions}
SKS and VT contributed equally by performing the computer simulations, collecting the data and developing analysis tools. SKS, VT and AS conceived and designed the analysis. All the authors performed the analysis and wrote the paper.

\section*{Conflicts of Interest}
There are no conflicts to declare

%\section*{\label{sec:acknwl}Acknowledgements}
\vspace{-0.2in}\section*{Acknowledgments}\vspace{-0.1in}
This work was supported by ANRF/ECRG/2024/002243/ENS and  IIT (ISM) Dhanbad FRS Project fund (Project no. FRS(204)/2023-2024/CHEMICAL). The authors acknowledge National Supercomputing Mission (NSM) for providing computing resources for the Param Himalaya HPC System, which is implemented by C-DAC and supported by the Ministry of Electronics and Information Technology (MeitY) and Department of Science and Technology (DST), Government of India.

\linespread{1.5}\selectfont
%%%%%%%%%%%%%%%%%%%%%%%%%%%%%%%%%%%%%%%%%%%%%%%%%%%%%%%%%%%%%%%%%%%%%
\section*{\label{sec:Appendix}Appendix: Velocity correlation function}
%%%%%%%%%%%%%%%%%%%%%%%%%%%%%%%%%%%%%%%%%%%%%%%%%%%%%%%%%%%%%%%%%%%%%

\begin{figure*}[ptbh]
  \centerline{
 \resizebox{\textwidth}{!}{ \begin{tabular}{cc}
        \includegraphics[width=15cm,height=!]{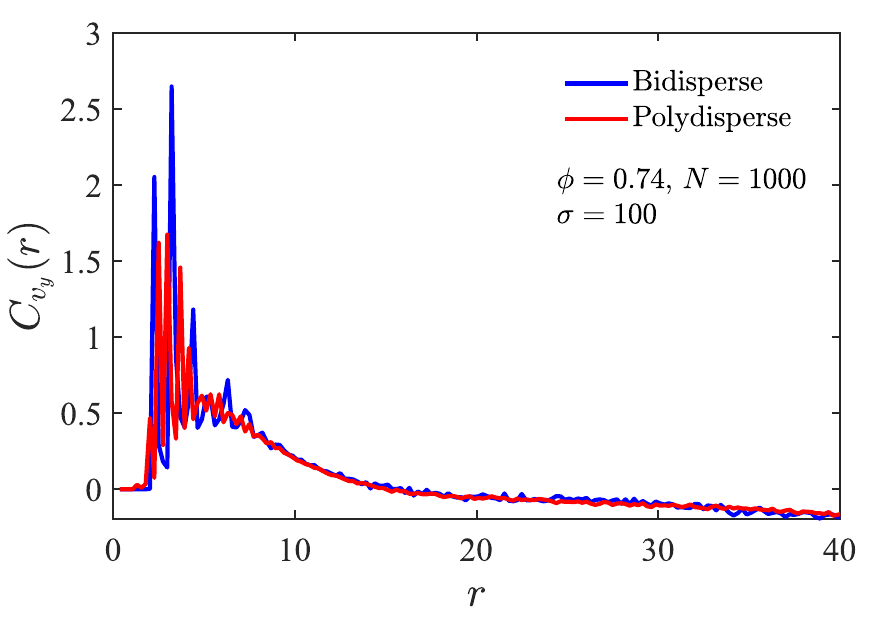} 
    &
        \includegraphics[width=15cm,height=!]{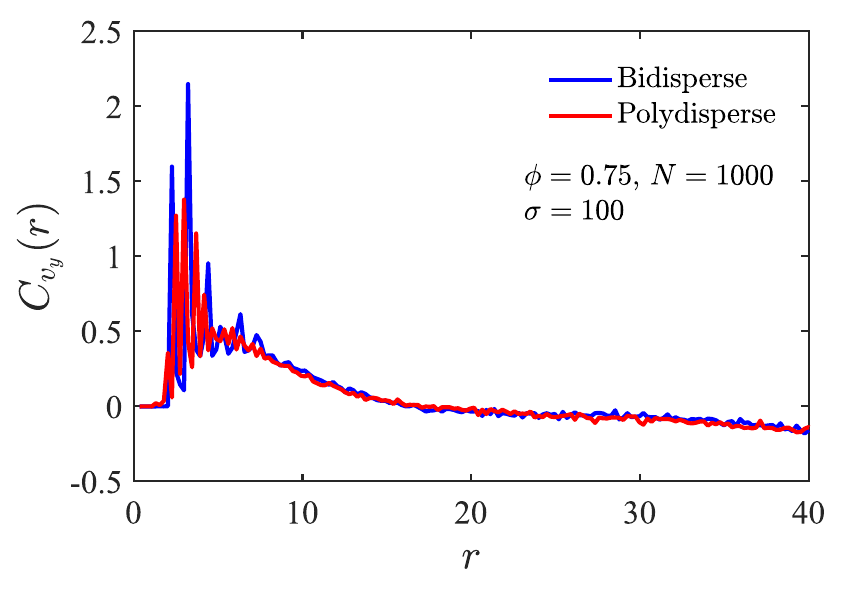}  \\[5pt]
         (a)  & 
       (b) \\
 \end{tabular} }}   
\caption{\small{Comparison of gradient direction velocity correlation function $C_{v_y}(r)$ of polydisperse and statistically equivalent bidisperse suspensions for $\alpha=0.133$, $N=1000$ and $\sigma=100$, at packing fractions (a) $\phi=0.74$ and (b) $\phi=0.75$.} 
\label{fig:vel_corr}}
\end{figure*}

The equivalence in the microstructure of polydisperse suspensions with that of the statistically equivalent bidisperse suspensions is presented here by comparing the gradient direction velocity ($v_y$) correlation function, defined as,

\begin{equation}
    C_{v_y}(r) = \langle\frac{1}{N}\sum\limits_{i<j} v_{y_i}\,v_{y_j}\,\delta(r-r_{ij})\rangle 
\end{equation}
where, for a given value of radial distance $r$, $C_{v_y}(r)$ is averaged over all the simulation snapshots. As shown in Figs.~\ref{fig:vel_corr} (a) and (b), the velocity correlation functions show similar variation for the polydisperse and statistically equivalent bidisperse suspensions. Interestingly, it is noted that at short distances (for smaller $r$) the fluctuations in the correlation function for polydisperse case is not exactly overlapping with the equivalent bidisperse systems. This is because in bidisperse systems there are only three types of neighbouring contacts based on particle size, \textit{i.e.}, \textit{small-small}, \textit{large-large} and \textit{large-small}, whereas, for the polydisperse systems there are more types of near-neighbour contacts, resulting in a different contact distribution at small $r$ for polydisperse systems as compared to equivalent bidisperse suspensions. However, this difference in fluctuation of the correlation function gets minimized with increasing radial distance (at long range). Thus, except for a deviation at short range, the motion correlation of polydisperse and equivalent bidisperse suspensions are almost identical.

%\mciteErrorOnUnknownfalse
\bibliography{bibfile}
\bibliographystyle{unsrtnat}

\end{document}

%% file: operands.tex
\usepackage{amsmath}

\DeclareMathAlphabet{\mathsf}{OT1}{phv}{b}{n}

\newcommand{\crossVorg}{\ensuremath{%
         \setbox0=\hbox{$V$}
        V \kern-\wd0{\raise.3ex\hbox{$\relbar$}}}}

\newcommand{\crossVxx}[2]{%
	{\setbox0=\hbox{$#1#2V$}
         \setbox1=\hbox{$#1#2$}
         \setbox2=\hbox{$#1V$}
         \dimen1=\wd0
	 \advance\dimen1-\wd1
         \raise.2\ht0\hbox{$#1#2$}\kern-.4\wd0}}

\usepackage{pifont}